\documentclass[twocolumn,aps,prd,amsmath,amssymb,preprintnumbers,longbibliography]{revtex4-1}
\usepackage{amsmath} \usepackage{amsfonts} \usepackage{amssymb}
\usepackage{bbm}
\usepackage{epsfig}
\usepackage{graphics}
\usepackage{graphicx}
\usepackage{titlesec}
\usepackage{mathtools}
\usepackage{environ}
\usepackage{dsfont}

\usepackage[colorlinks=true]{hyperref}
\hypersetup{urlcolor=black,linkcolor=black,citecolor=black}
\usepackage[toc,page]{appendix}

\makeatletter
\renewcommand*\env@matrix[1][c]{\hskip -\arraycolsep
  \let\@ifnextchar\new@ifnextchar
  \array{*\c@MaxMatrixCols #1}}
\makeatother

\titleformat{\subsection}[block]{\normalfont\bfseries}{\thesubsection.}{1ex}{}
\titlespacing{\subsection}{0pt}{10pt}{1pt}[0pt]
\titleformat*{\section}{\large\bfseries}
\renewcommand{\thesubsection}{\arabic{subsection}}
\numberwithin{equation}{section}

\usepackage[utf8]{inputenc}
\usepackage[T1]{fontenc}
\usepackage[svgnames]{xcolor}
\usepackage{microtype,enumitem,lmodern,dsfont}

\usepackage{mathtools,amssymb,tensor}

\usepackage{cleveref}
\Crefname{equation}{Eq.}{Eqs.}
\crefname{section}{sec.}{secs.}
\crefname{appendix}{app.}{apps.}
\crefname{cond}{condition}{conditions}
\AtBeginDocument{}
\newcounter{condition}[equation]

\DeclareMathOperator{\tr}{tr}
\DeclareMathOperator{\Tr}{Tr}
\DeclareMathOperator{\Det}{Det}
\newcommand{\transp}{\mathrm{T}}

\newenvironment{alignedeqn}{\begin{equation}\begin{aligned}}{\end{aligned}\end{equation}\ignorespacesafterend}

\begin{document}
\unitlength=1mm 

\title[ ]{Gauge-invariant fields and flow equations for Yang-Mills theories}

\author{C. Wetterich}

\affiliation{\href{eqn:mailto:c.wetterich@thphys.uni-heidelberg.de}{c.wetterich@thphys.uni-heidelberg.de}\\
{Universität Heidelberg, Institut für Theoretische Physik, Philosophenweg 16, D-69120 Heidelberg}}

\begin{abstract}
We discuss the concept of gauge-invariant fields for non-abelian gauge theories. Infinitesimal fluctuations around a given gauge field can be split into physical and gauge fluctuations. Starting from some reference field the gauge-invariant fields are constructed by consecutively adding physical fluctuations. An arbitrary gauge field can be mapped to an associated gauge invariant field. An effective action that depends on gauge-invariant fields becomes a gauge-invariant functional of arbitrary gauge fields by associating to every gauge field the corresponding gauge-invariant field. The gauge-invariant effective action can be obtained from an implicit functional integral with a suitable ``physical gauge fixing''. We generalize this concept to the gauge-invariant effective average action or flowing action, which involves an infrared cutoff. It obeys a gauge-invariant functional flow equation. We demonstrate the use of this flow equation by a simple computation of the running gauge coupling and propagator in pure $SU(N)$-Yang-Mills theory.
\end{abstract}

\maketitle

\hypertarget{TOC}{\tableofcontents}

\section{Introduction}

In a quantum field theory context the field equations employed in ``classical field theory'' correspond to exact equations that are obtained by the first functional derivative of the quantum effective action. These field equations are used widely in practice. For electromagnetism in vacuum the lowest order are Maxwell's equations, while effects of quantum fluctuations are taken into account by the Euler-Heisenberg correction \cite{HE1,JS,HE2}\nocite{HE2} in the effective action. In condensed matter physics the field equations obtained from a suitable effective action (``Landau theory'') describe superconductivity and many other phenomena. For electromagnetism, a crucial ingredient is the gauge invariance of the effective action.

The field equations derived from the quantum effective action for gravity are the basis of general relativity and cosmology. Modifications of the Einstein-Hilbert action, for example by higher-order curvature terms, can describe the inflationary epoch in cosmology \cite{Sta}. Again, a central ingredient is diffeomorphism invariance (invariance under general coordinate transformations) of the effective action. Quantum gravity is characterized by a non-abelian local gauge symmetry.

The validity of the field equations for electromagnetism and gravity has been tested by numerous precision experiments. For Yang-Mills theories, such as quantum chromodynamics or the electroweak gauge theory, field equations are explored widely as well. For these theories the direct observational tests of the field equations are more difficult. Still, the Higgs phenomenon \cite{Hi,BE} is a direct consequence of the quantum field equations for the electroweak gauge theory coupled to scalar fields.

In practice, one is used to treat fields as observables. We measure electric and magnetic fields, or the metric field in gravity. At first sight, this seems to conflict with the observation that the metric changes under gauge transformations, and with the general property that only gauge invariant objects are observables. For local gauge theories formulated in terms of continuum gauge fields the status of the arguments of the effective action seems then to differ from simple scalar theories. For the latter the value of the scalar field may correspond to magnetization and is directly observable. For electromagnetism the issue has a simple solution. Transversal gauge fields are gauge invariant. They can be associated with physical gauge fields that are observable. Imposing the gauge $\partial^{\mu}A_{\mu}=0$ we can determine the transversal gauge fields that correspond to given configurations of electric and magnetic fields.

If one ``measures the metric'' one should do something similar. One again has to fix a gauge in order to eliminate the redundancy of the gauge transformations. The notion of transversal fields for electrodynamics has to be generalized to a construction of gauge invariant fields for gravity. This extends to other non-abelian gauge theories as Yang-Mills theories. Using gauge invariant fields as observables provides for a much more direct contact to practical measurement than any attempt to express the detailed outcome of a measurement of the metric field of the earth in terms of invariants as the curvature scalar or similar. In this paper we construct gauge invariant fields for non-abelian gauge theories. They will be the crucial concept for our construction of the gauge invariant effective action.

While it is rather clear that the gauge-invariant quantum effective action is a powerful concept, its construction in a quantum field theory for local gauge theories is less obvious. The reason is that perturbative approaches require gauge fixing, whereas for non-perturbative methods, such as lattice gauge theories, the implementation of the continuum gauge fields and their effective action is difficult. In a continuum formulation the gauge fixing is necessary since the second functional derivative of a gauge-invariant action contains zero modes -- they correspond to the gauge fluctuations. The propagator as a central ingredient for all computations of fluctuation effects is then not well-defined. The solution of this problem by gauge fixing guarantees a well-defined propagator, but the gauge invariance of the effective action is now lost.

Background gauge fixing \cite{GH,BDW,BB,LA} uses a ``background field'' $\bar{A}_{\mu}$ in the gauge fixing condition. This restores invariance of the effective action under simultaneous gauge transformations of the macroscopic gauge field $A_\mu$ and the background field $\bar{A}_\mu$. The gauge-invariant effective action depends now on two gauge fields $A_\mu$ and $\bar{A}_\mu$. This is, however, not the object used in electrodynamics or general relativity, where the gauge-invariant effective action depends only on one gauge field or metric. A gauge-invariant effective action depending on a single gauge field can be obtained by identifying $\bar{A} = A$. This object depends, however, on the gauge fixing and its connection to physical observables like the quantum field equations and correlation functions is not obvious.

In this work we propose a gauge-invariant effective action which depends on a single gauge field. The field equations derived from this effective action are the equations of ``classical field theory'' discussed in the beginning. Our continuum construction is similar to background gauge fixing in a particular ``physical gauge'', which can be taken as Landau gauge for Yang-Mills theories. We do not introduce a separate ``fixed'' background field $\bar{A}_\mu$. The gauge fixing is formulated with the macroscopic field $A_\mu$ replacing the usual background field $\bar{A}_\mu$, such that the effective action will depend on a single macroscopic gauge field. With this construction the functional integral defining the effective action is turned into an implicit construction, involving a functional differential equation.

As an example, we may define the gauge-invariant effective action $\bar\Gamma[A]$ by a functional integral over the fluctuating microscopic gauge field $A^\prime$,
\begin{alignedeqn}\label{eqn:AA1}
	\bar\Gamma[A]
	&= -\ln\biggl(\int \mathcal{D} A^\prime \, \prod_z \delta(G^z) \, \tilde{M}\\
	&\hphantom{{}=}\times \exp\biggl\{-S[A^\prime] + \int_x \frac{\partial \bar\Gamma}{\partial A_\mu^z} \, (A_\mu^{\prime z} - \hat{A}_\mu^z)\biggr\}\biggr),
\end{alignedeqn}
where
\begin{equation}\label{eqn:AA2}
	G^z
	= \bigl[D^{-2}(\hat{A}) \, D^\mu(\hat{A}) (A_\mu^\prime - \hat{A}_\mu)\bigr]^z,
\end{equation}
and
\begin{equation}\label{eqn:AA3}
	\delta(G^z)
	= \lim_{\alpha \to 0} \exp\biggl\{-\frac{1}{2 \alpha} \int_x (G^z)^2\biggr\}.
\end{equation}
Here we associate to every macroscopic gauge field $A$ a gauge invariant physical gauge field $\hat{A}(A)$. The effective action is evaluated for physical gauge fields, and subsequently extended to arbitrary gauge fields, $\bar{\Gamma}[A]=\bar{\Gamma}[\hat{A}(A)]$. Covariant derivatives $D^\mu(\hat{A})=D^{\mu}(\hat{A}(A))$ are taken with the physical gauge field, and $D^2 = D_\mu \, D^\mu$. The microscopic or classical action $S[A^\prime]$ for the microscopic gauge field $A^\prime$ is assumed to be gauge-invariant, and we omit field-independent normalization factors. The Faddeev-Popov determinant
\begin{equation}\label{eqn:3A}
	\tilde{M}
	= \bigl[\det\bigl(-D^2(\hat{A})\bigr)\bigr]^{-1} \, \det\bigl(-D^\mu(\hat{A}) \, D_\mu(A^\prime)\bigr)
\end{equation}
equals one for $A_\mu^\prime = \hat{A}_\mu$. With
\begin{align}\label{eqn:AA4}
	\prod_z \delta(G^z)
	&= \det\bigl(-D^2(\hat{A})\bigr) \lim_{\alpha \to 0} \prod_z\\
	&\hphantom{{}=} \times \exp\biggl\{-\frac{1}{2 \alpha} \int_x \Bigl(\bigl[D^\mu(\hat{A}) (A_\mu^\prime - \hat{A}_\mu)\bigr]^z\Bigr)^2\biggr\},\notag
\end{align}
the factor $\det(-D^2)$ cancels the first factor in $\tilde{M}$ and one recognizes the close resemblance to Landau gauge in the background field formalism. We typically work with (infinitesimally) small $\alpha$ and take the limit $\alpha \to 0$ at the end. Possible improvements of the physical gauge fixing and the precise functional integral definition of $\bar\Gamma[A]$ will be discussed later in the text and in the appendices.

While the definition \eqref{eqn:AA1} is conceptually a rather complicated object, there will never be a need for an explicit solution of this functional differential equation. A perturbative expansion of the gauge-invariant effective action, or a computation by non-perturbative functional renormalization, involves only techniques familiar from background gauge fixing. Our approach based on physical gauge fields involves a particular gauge fixing. It differs in this respect from gauge invariant formulations that employ a gauge invariant regularization \cite{MOR1,MOR2,MOR3}. The advantage is the simplicity of the setting, in practice close to existing computations.

A key element in our approach are the gauge invariant physical gauge fields $\hat{A}(A)$. Our construction of physical gauge fields $\hat{A}$ exploits the split of infinitesimal fluctuations $A_\mu^\prime - A_\mu$ into physical and gauge fluctuations. Gauge-invariant physical fields can be constructed by starting from some reference field $\bar{A}_r$, for example $\bar{A}_{r,\mu} = 0$, and consecutively adding physical fluctuations. Gauge-invariant fields obey differential constraints. We can also view the physical gauge fields $\hat{A}$ as the unique representations for every gauge orbit. In this way one associates to every gauge field $A$ a physical gauge field $\hat{A}(A)$. 

An arbitrary gauge field $A_\mu$ can be decomposed into a physical gauge field $\hat{A}_\mu(A)$ and a gauge degree of freedom $\hat{c}_\mu(A)$,
\begin{equation}
	A_\mu
	= \hat{A}_\mu + \hat{c}_\mu.
\end{equation}
This decomposition depends on the choice of the reference field $\bar{A}_r$. A gauge-invariant effective action only depends on the physical gauge fields $\hat{A}_\mu$ and is independent of $\hat{c}_\mu$. The distinction between physical and gauge degrees of freedom is essentially a local issue in field space, referring to infinitesimal changes of gauge fields and a definition by differential relations. It is this feature that necessitates the (arbitrary) choice of a reference field $\bar{A}_r$ for any global definition of gauge-invariant fields. A global definition of physical gauge fields not involving a reference field is possible only for abelian gauge theories.

The physical gauge fixing acts only on the gauge fluctuations, leaving the physical fluctuations untouched. As a result, the gauge-fixed effective action $\tilde\Gamma[A]$ becomes effectively a sum of two pieces $\tilde{\Gamma}=\bar{\Gamma}+\Gamma_{gf}$. The first is the effective action $\bar\Gamma[\hat{A}]$ for the physical gauge fields, while the second is a gauge-fixing term $\Gamma_{gf}[\hat{A},\hat{c}]$, which is quadratic in the macroscopic gauge fluctuations $\hat{c}$, with a coefficient tending to infinity. A partial solution of the field equations for $\hat{c}$ therefore implies $\hat{c}_\mu = 0$. Inserting this solution into $\tilde\Gamma$ eliminates the gauge-fixing term $\Gamma_\text{gf}$, such that only the effective action $\bar\Gamma[\hat{A}]$ remains. The gauge-invariant effective action $\bar\Gamma[A]$ is obtained by extension, associating to every $A$ the physical field $\hat{A}(A)$ and defining $\bar\Gamma[A] = \bar\Gamma[\hat{A}(A)].$ This realizes the general mechanism of how a gauge-invariant effective action can arise via the ``decoupling'' of gauge modes \cite{CWGD}. 

The first functional derivative of the gauge-invariant effective action yields a source term which is covariantly conserved
\begin{equation}\label{eqn:conservation law}
	\frac{\partial \bar\Gamma}{\partial A_\mu}
	= J^\mu,
	\qquad
	D_\mu J^\mu
	= 0.
\end{equation}
Eq.~\eqref{eqn:conservation law} constitutes in a quantum field theory context the ``classical field equations'' in the presence of sources. As it should be, these field equations do not involve the microscopic or ``classical'' action, but the effective action which includes all fluctuation effects. In a quantum field theory the relevant source term $J^{\mu}$ is defined by the quantum field equation \eqref{eqn:conservation law}. It may differ from some microscopic source term. In quantum gravity the r.h.s. of the quantum field equation involves the effective or renormalized energy momentum tensor.

The second functional derivative $\bar\Gamma^{(2)}$ of the gauge-invariant effective action constitutes the inverse propagator for the physical fluctuations. On the space of all fluctuations only $\tilde\Gamma^{(2)}$ is invertible, while $\bar\Gamma^{(2)}$ contains zero modes corresponding to gauge fluctuations. We can project the fluctuations on the physical fluctuations employing a projector $P$. Then $\bar\Gamma^{(2)}$ becomes invertible on the space of physical fluctuations
\begin{equation}\label{eq:AS:8} 
	\bar\Gamma^{(2)} \, G
	= P^\transp,
\end{equation}
with $G$ the propagator (Green's function) for the physical fluctuations. We discuss the relation of $G$ to the connected two-point correlation function for the physical fluctuations. Knowledge of the gauge-invariant effective action $\bar\Gamma$ is sufficient to compute the correlation function for the physical fluctuations. One-particle-irreducible higher-order correlation functions for the physical fluctuations are obtained from higher functional derivatives of $\bar\Gamma$.

Our construction shows some similarities with the geometric formulation of gauge theories by Vilkovisky and de Witt \cite{Vil,DeW,REB,PG1,BR,DE}. This concerns the property that the macroscopic field does not equal, in general, the expectation value of the microscopic field. There is also a common emphasis on gauge orbits and their representatives. Our main emphasis concerns the notion of gauge invariant physical gauge fields which is not present in the Vilkovisky-DeWitt formulation. They are the basis of our construction of a gauge invariant effective action. On the other hand, the gauge invariant effective action proposed here is not parameterization invariant, which is the main concept in ref.~\cite{Vil,DeW,REB,PG1,BR,DE}. The formulation \eqref{eqn:AA1} involves explicitly $A^{\prime}$, which is a ``coordinate'' in configuration space. Also the construction of $\hat{A}$ is not parameterization invariant. Formally, the two approaches are therefore rather different, but it seems not excluded that some new relations may be found. This may concern, for example, an explicit construction for the physical fields $\hat{A}(A)$ by employing the connection in configuration space used in ref.~\cite{KUN,DWMP}.

Our setting can be extended to a gauge-invariant effective average action or flowing action $\Gamma_k[A]$, for which only fluctuations with covariant momenta larger than $k$ are included (in a Euclidean setting). Following ref.~\cite{CWGIF} we will discuss a gauge-invariant flow equation for the effective average action $\Gamma_k[A]$,
\begin{equation}\label{eqn:I1}
	k \partial_k \Gamma_k[A]
	= \frac{1}{2} \tr\bigl\{k \partial_k R_P(A) \, G_P(A)\bigr\} - \delta_k[A].
\end{equation}
Here $G_P$ is the propagator for the physical fluctuations in the background of an arbitrary macroscopic gauge field $A_\mu$ and in presence of the infrared cutoff. It generalizes $G$ in eq.~\eqref{eq:AS:8} for $k\neq 0$. The infrared cutoff function $R_{P}$ suppresses fluctuations with covariant momenta smaller than $k$ and renders the momentum integral contained in the trace in eq.~\eqref{eqn:I1} infrared finite. The term $k\partial_{k}R_{P}$ vanishes fast for high momenta, ensuring that eq.~\eqref{eqn:I1} is also ultraviolet finite. (In the following we often will omit the label $k$ for $\Gamma_k[A]$.) For any setting where $G$ and $R$ are formulated in an extended field space including gauge fluctuations one can employ a projection on the physical fluctuations,
\begin{alignedeqn}\label{eqn:I2}
	&G_P(A)
	= P(A) \, G(A) \, P^\transp(A),\\
	&R_P(A)
	= P^\transp(A) \, R(A) \, P(A).
\end{alignedeqn}
More generally, $G_{P}$ obeys $PG_{P}=G_{P}P^{T}=G_{P}$, and similar for $R_{P}$.

For pure Yang-Mills theories the projector onto physical fluctuations $P$, $P^2 = P$, is given by
\begin{equation}\label{eqn:228A}
	\tensor{P}{_\mu^\nu}
	= \delta^\nu_\mu-D_\mu D^{-2}D^\nu.
\end{equation}
It involves covariant derivatives $D_\mu$ formed with the macroscopic gauge field $A_\mu$, with $D^2 = D_\mu D^\mu$. The propagator $G_P$ is related to the projected second functional derivative of $\Gamma$,
\begin{alignedeqn}\label{eqn:I5}
	\Gamma^{(2)}_P
	&= P^\transp \, \Gamma^{(2)} \, P,\\
	\Gamma^{(2)\mu\nu}_{zy}(x,y)
	&= \frac{\partial^2\Gamma}{\partial A^z_\mu(x)\partial A^y_\nu(y)}.
\end{alignedeqn}
It involves the infrared cutoff according to
\begin{equation}\label{eqn:I6}
	\bigl(\Gamma_P^{(2)} + R_P\bigr) G_P
	= P^\transp.
\end{equation}
While $\Gamma^{(2)}$ is not invertible due to the presence of zero modes associated to gauge invariance, the inverse propagator $\Gamma_P$ for the physical fluctuations is invertible on the appropriate projected subspace.

The infrared cutoff function $R$ depends on the renormalization scale $k$. It typically involves covariant derivatives formed with the macroscopic gauge field $A_\mu$. The measure contribution $\delta_k(A)$ arises from the regularization of the factor $\prod_z \delta(G^z) \, \tilde{M}$ in eq.~\eqref{eqn:AA1} or, more generally, from the regularization of the gauge modes and the Faddeev-Popov determinant. It is a given function of covariant derivatives, typically depending on $D^2(A)/k^2$. It does not involve the effective action $\Gamma[A]$ and its functional derivatives. The flow equation \eqref{eqn:I1} is closed in the sense that for any macroscopic gauge field $A$ the r.h.s. can be evaluated in terms of $\Gamma^{(2)}$. No separate background field is involved.

For $k = 0$ the effective average action $\Gamma_k[A]$ equals the gauge-invariant quantum effective action $\bar\Gamma[A]$ -- all fluctuations are included. On the other hand, for $k \to \infty$, or $k$ equal to some large UV-scale $\Lambda_\text{UV}$, no fluctuations are included. In this region of very large $k$ the effective average action $\Gamma_k$ equals the microscopic or ``classical'' action. The solution of the flow equation interpolates from the microscopic action for large $k$ to the macroscopic or quantum effective action for $k \to 0$.

The precise status of the gauge invariant flow equation \eqref{eqn:I1}, if it is exact or only a good approximation, depends on the choice of the relation between the macroscopic gauge field $A$ and the expectation value of the microscopic gauge field $\langle A^{\prime}\rangle$. Only for an ``optimal choice'' of this relation eq.~\eqref{eqn:I1} becomes exact, which requires the existence of a solution of a differential relation \cite{CWGIF}. In general, there exists always a closed gauge invariant flow equation, but it may involve additional terms if the choice of $A(\langle A^{\prime}\rangle )$ is not optimal. Since this is not the main emphasis of the present paper we discuss the issue in an appendix \ref{app:gauge-invariance flow}. We also note that the flow equation is manifestly gauge invariant, while the regularization is not, since $S_{gf}$ involves the gauge fluctuations. This differs from the construction of explicitly gauge invariant regulators in refs.~\cite{MOR1,MOR2,MOR2}. The issue of dependence of the flow equation on the choice of gauge \cite{LAV} is, at least partly, settled by the restriction to a physical gauge.

In practice, the flow equation \eqref{eqn:I1} resembles closely the flow equation in the background formalism \cite{RW} in Landau gauge, with background field $\bar{A}$ identified with $A$. It omits, however, the ``correction terms'' discussed in ref.~\cite{RW}, and employs a particular regularization of the Faddeev-Popov determinant. It can be seen as an a posteriori justification for the omission of the correction terms in many past practical computations, by use of a different object that follows a different flow trajectory in the ``theory space'' of functionals. Past investigations of functional renormalization in Landau gauge \cite{RW,R2,R3,R4} have computed the running of the gauge coupling in various dimensions \cite{RW,R3,R11,R12,R13} or at different temperatures \cite{RW,R14}. They have addressed the phase diagram of superconductors \cite{R6,R7}, gluon condensation \cite{R18,R19}, the heavy quark potential \cite{R15,R16,R17,EHW}, the gluon propagator \cite{R16,R19,R20,R23,R24,FI,CMPS1,CMPS2}, as well as confinement \cite{R21} and the infrared properties of QCD \cite{R23,R24,R25}, see refs.~\cite{R8,R9,R10} for reviews.

There are also important practical differences to the background formalism in Landau gauge. Since the flowing action is gauge-invariant and depends only on one macroscopic field, there is no need to investigate modified Slavnov-Taylor identities or background identities \cite{R53,R54,R55,R56,R57,R58,R59}. Differentiation of the flow equation \eqref{eqn:I1} with respect to the macroscopic gauge field $A_\mu$ commutes with the $k$-derivative. Concerning the flow of $n$-point functions the flow of the proposed gauge-invariant flowing action involves additional diagrams as compared to the background field formalism. They are generated by $A_\mu$-derivatives of $R_P$ in eq.~\eqref{eqn:I1}, and similarly by the $A_\mu$-dependence of the IR cutoff in the measure term $\delta_k$. On the other hand, it is an important technical simplification that the flow of $n$-point functions can now directly be found by functional derivatives of the gauge-invariant flow generator in eq.~\eqref{eqn:I1}. This is not possible in the usual background field formalism. We also recall that even for physical gauge fixing the effective action with background field $\bar{A} = A$ differs from the one with $\bar{A} = 0$. As one of the important effects the wave function renormalization for $A$ differs \cite{R17}.

An important issue for the practical usefulness of the gauge invariant effective action and the gauge invariant flow equation is the question of locality. The quantum effective action is not a local object in the strict sense. Already the perturbative running of the gauge coupling induces a logarithmic dependence on momentum that cannot be described by any finite polynomial. In the non-perturbative range the effective gluon propagator may involve a non-local mass term. These types of non-localities express physical properties and cannot be avoided. The question arises if additional ``spurious'' non-localities are generated by our formalism, since the latter involves non-local projectors. In the computations performed so far no such ``spurious'' non-localities have shown up. The reason is mainly that the projections can be implemented indirectly by a physical Landau gauge gauge fixing, and the latter has a local nature. Presumably, only practical experience will finally settle this issue.

This paper is organized as follows. In sect.~\ref{sec:gauge fields sources} we recapitulate the general connection between the effective action and sources. Particular emphasis is paid to physical sources obeying the conservation law \eqref{eqn:conservation law}. The projector onto the conserved sources is the same as the one projecting onto physical fluctuations of the macroscopic gauge field. This establishes the close connection between physical fluctuations and physical sources. In sect.~\ref{sec:physical gauge fields} we introduce the notion of gauge-invariant fields or physical gauge fields.

In sect.~\ref{sec:functional integral} we construct the gauge-invariant effective action from a functional integral and compare it to the background formalism. Sect.~\ref{sec:gauge-invariant flow} turns to the flow equation for the gauge-invariant effective action \eqref{eqn:I1}. We define the measure contribution $\delta_k$. As a practical demonstration we compute the running gauge coupling and the flow of the propagator in $SU(N)$-Yang-Mills theory. Sect.~\ref{sec:conclusions} contains our conclusions. In a series of appendices we discuss a non-linear formulation of the split into physical gauge-invariant fields and gauge degrees of freedom, as well as the general consequences of a setting where the macroscopic field appears in the gauge fixing for the microscopic field.

\section{Gauge fields and sources}
\label{sec:gauge fields sources}

In this section we discuss the split of sources and infinitesimal fluctuations of gauge fields into a physical part and a gauge part. We introduce the appropriate projectors.

\subsection{Gauge transformations}

For the discussion of gauge transformations we employ matrix valued gauge fields $A_\mu\left( x \right) = A_\mu^z t_z$, where $t_z$ are the generators of the gauge group in the fundamental representation. For an abelian U(1)-gauge group one has $t = 1/\sqrt{2}$. For $SU(2)$ the generators are given by the Pauli matrices, $t_z = \tau_z/2$. We normalize the generators as
\begin{equation}\label{eqn:GA}
	\Tr(t_y t_z)
	= \frac{1}{2} \, \delta_{yz},
\end{equation}
where $\Tr$ stands for the trace over internal indices. For the field variables $A^z_\mu$ we use a normalization for which the covariant derivative in the fundamental representation reads
\begin{equation}\label{eqn:GB}
	D_\mu
	= \partial_\mu - i A_\mu,
	\qquad
	A_\mu
	= A^z_\mu t_z.
\end{equation}

Infinitesimal gauge transformations act on gauge fields as
\begin{equation}\label{eqn:G1}
	\delta A_\mu
	= D_\mu \varphi
	= \partial_\mu \varphi - i [A_\mu,\varphi],
	\qquad
	\varphi
	= \varphi^z t_z.
\end{equation}
The field strength,
\begin{equation}\label{eqn:G2}
	F_{\mu\nu}
	= \partial_\mu A_\nu - \partial_\nu A_\mu - i [A_\mu,A_\nu],
\end{equation}
transforms as
\begin{equation}\label{eqn:G3}
	\delta F_{\mu\nu}
	= i [\varphi,F_{\mu\nu}],
\end{equation}
such that $\Tr F^{\mu\nu} F_{\mu\nu} = \frac{1}{2} \, F^z_{\mu\nu}F_z^{\mu\nu}$ is invariant.

Generating functionals for correlation functions are obtained by adding to the microscopic action a source term. The partition function $Z[L]$ is defined by the functional integral
\begin{equation}
	Z[L]
	= \int \mathcal{D} A^\prime \, \exp\bigl\{-\tilde{S}[A^\prime] - S_L[A^\prime,L]\bigr\},
\end{equation}
where $\tilde{S}$ includes gauge fixing and the Faddeev-Popov determinant. On the microscopic level an action for the source $L^\mu$ typically involves two pieces
\begin{equation}\label{eqn:J1}
	S_L
	= -2 \int_x \Tr(L^\mu A_\mu^\prime) + \int_x \tilde{L}.
\end{equation}
The first is the generic source term, while the second may be needed in order to guarantee the gauge invariance of $S_L$. If the two pieces transform under infinitesimal gauge transformations as
\begin{equation}\label{eqn:J2}
	\delta L^\mu
	= i [\varphi,L^\mu],
	\qquad
	\delta \tilde{L}
	= 2 \Tr(\partial_\mu \varphi L^\mu),
\end{equation}
the source term $S_L$ is gauge invariant. For the example of gauge fields coupling to fermions the covariant fermion kinetic term
\begin{equation}\label{eqn:J3}
	S_L
	= i \int_x \bar\psi \gamma^\mu (\partial_\mu - i A^{\prime z} t_z) \psi
\end{equation}
amounts to
\begin{equation}\label{eqn:J4}
	L^\mu_z
	= -\bar \psi\gamma^\mu t_z\psi,
	\qquad
	\tilde{L}
	= i\bar \psi\gamma^\mu\partial_\mu\psi.
\end{equation}
With
\begin{equation}\label{eqn:J5}
	\delta\psi
	= i \varphi \psi
	= i \varphi^z t_z \psi,
	\qquad
	\delta\bar\psi
	= -i \bar\psi \varphi
\end{equation}
one indeed finds the transformation property \eqref{eqn:J2}. The term $\sim \tilde{L}$ does not involve $A^\prime$ or $A$ and yields for $W = \ln Z$ a simple additive ``constant'' $-\int_x \tilde{L}$.

\subsection{Projectors}

Physical sources are covariantly conserved, obeying the constraint
\begin{equation}\label{eqn:G9}
	D_\mu J^\mu
	= \partial_\mu J^\mu - i [A_\mu,J^\mu]
	= 0.
\end{equation}
For general sources $L^\mu$ the projection on the physical sources obeys
\begin{equation}\label{eqn:G8}
	J^\mu
	= \tensor{P}{^\mu_\nu} L^\nu
	= L^\mu - D^\mu D^{-2} D_\nu L^\nu
	= L^\nu \, \tensor{(P^\transp)}{_\nu^\mu},
\end{equation}
where the projector $P$ is defined by
\begin{alignedeqn}\label{eqn:I4}
	&\tensor{P}{_\mu^\nu}
	= \tensor{\delta}{_\mu^\nu} - \tensor{\bar{P}}{_\mu^\nu},\\
	&\tensor{\bar{P}}{_\mu^\nu}
	= D_\mu D^{-2} D^\nu.
\end{alignedeqn}
This projector is a central object for our discussion.

The transposed projector obeys
\begin{equation}\label{eqn:228B}
	\tensor{(P^\transp)}{_\mu^\nu}
	= \tensor{P}{^\nu_\mu}
	= \eta^{\nu\rho} \tensor{P}{_\rho^\tau} \eta_{\tau\mu},
	\quad
	\tensor{(P^\transp)}{^\mu_\nu}
	= \tensor{P}{_\nu^\mu}.
\end{equation}
such that for Minkowski signature the difference between $P^\transp$ and $P$ is only a question of raising and lowering indices, $\tensor{(P^\transp)}{_\mu^\nu} = \tensor{P}{_\mu^\nu}$. We observe the identities
\begin{equation}
	D^\mu \tensor{P}{_\mu^\nu}
	= 0,
	\qquad
	\tensor{P}{_\mu^\nu} D_\nu
	= 0.
\end{equation}
Longitudinal fields are annihilated by the projector $P$,
\begin{equation}\label{eqn:G7}
	\tensor{P}{_\mu^\nu} D_\nu B
	= 0.
\end{equation}

For fields $B$ in the adjoint representation the covariant derivatives are formed for the matrix representation \eqref{eqn:GB} with the macroscopic field $A_\mu$ appearing in commutators, e.g.
\begin{equation}\label{eqn:225A}
	D_\mu(A) B_\nu
	= \partial_\mu B_\nu - i [A_\mu,B_\nu],
	\quad
	D^2 = D^\mu D_\mu.
\end{equation}
For non-abelian gauge theories $P$ depends on $A_\mu$. In contrast, for abelian gauge theories one has $D_\mu B_\nu =\partial_\mu B_\nu$, such that the projector is field-independent. If we employ $A_\mu = A_\mu^z \, t_z$, and similarly for $B_\mu$, the operator $D_\mu$ does not act as a simple matrix multiplication. This has to be taken into account for the notion of products as used in the definition of the projector in eq.~\eqref{eqn:G8}, \eqref{eqn:I4}.

For a discussion of projectors it is often more convenient to use a representation where the action of covariant derivatives can be viewed as matrix multiplication. This holds if we represent gauge fields and sources as vectors with components labeled by $z$. Indeed, for fields $B^z$ in the adjoint representation, such as the gauge fields, we can represent $D^\mu$ as a matrix multiplication
\begin{equation}\label{eqn:26A}
	\bigl(D_\mu(A) \, B\bigr)^z
	= \tensor{\bigl(D_\mu(A)\bigr)}{^z_w} \, B^w,
\end{equation}
with
\begin{equation}\label{eqn:23AA}
	\tensor{(D_\mu(A))}{^z_w}
	= \partial_\mu \delta_w^z - A_\mu^y \tensor{f}{_y^z_w},
\end{equation}
and $f_{yzw}$ the totally antisymmetric structure constants of the gauge group. Compatibility with the matrix representation $B = B^z \, t_z$ follows for arbitrary representations of the generators $t_z$ from
\begin{alignedeqn}\label{eqn:26B}
	(D_\mu \, B)_{vw}
	&= \bigl(\partial_\mu B - i [A_\mu,B]\bigr)_{vw}\\
	&= \tensor{(D_\mu)}{^z_y} \, B^y (t_z)_{vw}
	= (D_\mu B)^z (t_z)_{vw}.
\end{alignedeqn}
Eq.~\eqref{eqn:26A}, \eqref{eqn:23AA} yield the projector acting directly on the $(\mu,z)$ index pair
\begin{equation}\label{eqn:23AB}
	P_{\mu y}^{z \nu}
	= \delta_\mu^\nu \, \delta_z^y - \tensor{(D_\mu)}{^z_w} \tensor{(D^{-2})}{^w_v}\tensor{(D^\nu)}{^v_y}.
\end{equation}
In this representation products of covariant derivatives and projectors are simply products of matrices containing differential operators. This facilitates many operations.

For arbitrary vectors $A$, $B$ partial integration yields for the scalar product
\begin{equation}
	\int_x B_z^\mu \tensor{\bigl(D_\mu \, D^{-2} \, D^\nu\bigr)}{^z_y} \, A_\nu^y
	= \int_x A_\nu^y \tensor{\bigl(D^\nu \, D^{-2} \, D_\mu\bigr)}{_y^z} \, B_z^\mu,
\end{equation}
justifying eq.~\eqref{eqn:228B}. According to practical convenience we will switch between the representations of gauge fields and sources by vectors or matrices. The use of vectors is typically indicated by the explicit index $z$, while for matrices we often do not indicate explicit indices.

\subsection{Physical sources and fluctuations}

If the constraint \eqref{eqn:G9} is realized for the sources appearing in the field equations derived from $\bar\Gamma[A]$,
\begin{equation}\label{eqn:G10}
	\frac{\partial\bar\Gamma}{\partial A^z_\mu}
	= {J^\mu_z},
	\qquad
	\frac{\partial\bar\Gamma}{\partial A_\mu}
	= J^\mu
	= J^\mu_z t_z,
\end{equation}
one concludes that $\bar\Gamma$ is gauge invariant,
\begin{align}\label{eqn:G11}
	\delta\bar\Gamma
	&= \int_x \frac{\partial\bar\Gamma}{\partial A^z_\mu}\delta A^z_\mu
	= 2 \int_x \Tr\left\{\frac{\partial\bar\Gamma}{\partial A_\mu}\delta A_\mu\right\}\\
	&= 2 \int_x \Tr\{J^\mu D_\mu\varphi\}
	= -2 \int_x \Tr\bigl\{D_\mu J^\mu)\varphi\bigr\}
	= 0.\notag
\end{align}
In other words, if the first derivative of $\bar{\Gamma}$ obeys
\begin{equation}\label{eqn:301A}
	D_\mu \frac{\partial\bar{\Gamma}}{\partial A_\mu}
	= 0,
\end{equation}
the source is conserved and $\bar{\Gamma}$ gauge invariant. We will realize the property \eqref{eqn:301A} by projecting a more general effective action $\Gamma$ on physical gauge fields.

With respect to the gauge transformations \eqref{eqn:G1} the physical sources transform homogeneously,
\begin{equation}\label{eqn:G11A}
	\delta J^\mu
	= i [\varphi,J^\mu].
\end{equation}
For $D_\mu J^\mu = 0$ the transformed physical source $J^{\prime\mu} = J^\mu + \delta J^\mu$ obeys $D^\prime_\mu J^{\prime\mu} = 0$. This can be seen from
\begin{align}\label{eqn:G11B}
	&D^\prime_\mu J^{\mu^\prime} - D_\mu J^\mu\\
	&= \partial_\mu J^{\prime\mu} - i [A_\mu + D_\mu\varphi, J^{\prime\mu}] - \partial_\mu J^\mu + i[A_\mu,J^\mu]\notag\\
	&= D_\mu\delta J^\mu - i[D_\mu\varphi,J^\mu]
	= i [\varphi,D_\mu J^\mu]
	= 0.\notag
\end{align}

We next turn to the physical fluctuations of gauge fields. The infinitesimal difference between two gauge fields,
\begin{equation}\label{eqn:J7}
	h_\mu
	= A^{(2)}_\mu - A^{(1)}_\mu,
\end{equation}
transforms homogeneously under simultaneous gauge transformations of $A^{(1)}$ and $A^{(2)}$.
\begin{equation}\label{eqn:J8}
	\delta h_\mu
	= i [\varphi,h_\mu].
\end{equation}
We can split
\begin{alignedeqn}\label{eqn:J9a}
	&h_\mu
	= f_\mu + a_\mu,
	&&f_\mu
	= \tensor{P}{_\mu^\nu} h_\nu,
	&&D^\mu f_\mu
	= 0,\\
	&a_\mu
	= D_\mu \lambda,
	&&\lambda
	= D^{-2} D^\nu h_\nu.
\end{alignedeqn}
The projector onto the physical fluctuations $f^\mu$ is the same as the one projecting onto the physical sources $J^\mu$.

Both $f_\mu$ and $a_\mu$ transform homogeneously, if $A^{(1)}_\mu$ and $A^{(2)}_\mu$ transform both according to eq.~\eqref{eqn:G1}. The gauge transformation of $A_\mu^{(1)} + h_\mu$ can equivalently be described by an inhomogeneous transformation of $a_\mu$ at fixed $A_\mu^{(1)}$,
\begin{equation}\label{eqn:J10}
	\delta f_\mu
	= i [\varphi,f_\mu],
	\qquad
	\delta a_\mu
	= D_\mu \varphi + i [\varphi,a_\mu].
\end{equation}
For infinitesimal $h_\mu$ we identify $a_\mu$ with the gauge fluctuations, while $f_\mu$ are the physical fluctuations. For both infinitesimal $h_\mu$ and $\varphi$ the gauge transformation only acts on the gauge fluctuations, $\delta h_\mu = \delta a_\mu = D_\mu \varphi$.

\section{Physical gauge fields}
\label{sec:physical gauge fields}

In this section we introduce the notion of gauge-invariant fields. For abelian gauge theories this can be implemented by a global constraint -- the gauge-invariant field $\hat{A}_\mu(x)$ is simply the transversal part of $A_\mu$, e.g. $\hat{A}_\mu = \tensor{P}{_\mu^\nu} A_\nu$. For non-abelian gauge theories a global constraint is no longer possible without the choice of a reference field. We rather realize a gauge-invariant field $\hat{A}_\mu(x)$ by imposing differential constraints. Starting from a reference field $\bar{A}_{r,\mu}(x)$ the gauge-invariant field is constructed by adding consecutively physical fluctuations \cite{CWGD}. The precise choice of the gauge-invariant field depends on the choice of the reference field.

\subsection{Gauge-invariant fields}

Following ref.~\cite{CWGD}, an arbitrary gauge field $A_\mu$ can be decomposed into a ``physical gauge field'' $\hat{A}_\mu$ and a gauge degree of freedom $\hat{c}_\mu$,
\begin{equation}\label{eqn:274A}
	A_\mu
	= \hat{A}_\mu + \hat{c}_\mu.
\end{equation}
The ``physical gauge fields'' $\hat{A}_\mu$ obey local differential constraints. Let two neighboring physical gauge fields differ by an infinitesimal fluctuation $\hat{h}_\mu,\hat{A}_\mu^{(2)} - \hat{A}^{(1)}_\mu = \hat{h}_\mu$. For infinitesimal $\hat{h}$ this fluctuation is required to be a physical fluctuation
\begin{equation}\label{eqn:G11C}
	\tensor{P}{_\mu^\nu}(\hat{A}) \, \hat{h}_\nu
	= \hat{h}_\mu,
	\qquad
	D^\mu \hat{h}_\mu
	= 0,
	\qquad
	\hat{h}_\mu
	= f_\mu.
\end{equation}
Once the notion of physical infinitesimal fluctuations $\hat{h}_\mu = f_\mu$ is established, the family of physical gauge fields $\hat{A}_\mu(x)$ can be constructed by starting with some field $\bar{A}_{r,\mu}(x)$, and then adding consecutively transversal or ``physical'' infinitesimal fluctuations $f_\mu$.

This construction can be cast into the form of differential constraints. Consider the change of $\hat{A}_\mu^z$ induced by an infinitesimal change of $A_\nu^y$
\begin{equation}\label{eqn:CA}
	\delta \hat{A}_\mu^z
	= \frac{\partial \hat{A}_\mu^z}{\partial A_\nu^y} \, \delta A_\nu^y.
\end{equation}
The difference between $\hat{A} + \delta \hat{A}$ and $\hat{A}$ is a physical fluctuation, resulting in the constraint
\begin{equation}\label{eqn:275A}
	\tensor*{P}{_\mu_w^z^\rho}(\hat{A}) \, \frac{\partial\hat{A}_\rho^w}{\partial A_\nu^y}
	= \frac{\partial\hat{A}_\mu^z}{\partial A_\nu^y}.
\end{equation}
The projector $P(\hat{A})$ involves covariant derivatives formed with $\hat{A}$.

On the other hand, if $\delta A$ is a pure gauge transformation, the physical gauge field $\hat{A}$ remains unchanged and one has by construction $\delta \hat{A} = 0$, or
\begin{equation}\label{eqn:CB}
	\frac{\partial \hat{A}_\mu^z}{\partial A_\nu^y} \bigl(1 - P(A)\bigr)_{\nu w}^{y \rho} \, \delta A_\rho^w
	= 0.
\end{equation}
In short, a gauge transformation of $A$ does not change the associated physical field $\hat{A}(A)$. This is expressed by a second constraint
\begin{equation}\label{eqn:XYA}
	\frac{\partial \hat{A}_\mu^z}{\partial A_\rho^w} \, P_{\rho y}^{w \nu}(A)
	= \frac{\partial \hat{A}_\mu^z}{\partial A_\nu^y}.
\end{equation}
As it should be, eq.~\eqref{eqn:XYA} directly implies the gauge invariance of $\hat{A}_\mu$
\begin{alignedeqn}\label{eqn:XYB}
	\delta \hat{A}_\mu^z
	&= \frac{\partial \hat{A}_\mu^z}{\partial A_\nu^y} \, \delta A_\nu^y
	= \frac{\partial \hat{A}_\mu^z}{\partial A_\nu^y} \, \bigl(D_\nu \varphi\bigr)^y\\
	&= \frac{\partial \hat{A}_\mu^z}{\partial A_\rho^w} \, P_{\rho y}^{w \nu} \, \bigl(D_\nu \varphi\bigr)^y
	= 0.
\end{alignedeqn}
By virtue of the constraint \eqref{eqn:XYA} one finds for an arbitrary fluctuation $h_\nu$,
\begin{equation}\label{eqn:275B}
	\frac{\partial\hat{A}_\mu}{\partial A_\nu} h_\nu
	= \frac{\partial\hat{A}_\mu}{\partial A_\rho} \tensor{P}{_\rho^\nu}(A) h_\nu
	= \frac{\partial\hat{A}_\mu}{\partial A_\nu} f_\nu.
\end{equation}

Using the properties of the transposed projector the two constraints can also be written as
\begin{equation}\label{eqn:CC}
	\frac{\partial \hat{A}_\mu^z}{\partial A_\nu^y}
	= \frac{\partial \hat{A}_\rho^w}{\partial A_\nu^y} \, \bigl(P(\hat{A})\bigr)_{w \mu}^{\rho z}
\end{equation}
and
\begin{equation}\label{eqn:CD}
	\frac{\partial \hat{A}_\mu^z}{\partial A_\nu^y}
	= \bigl(P(A)\bigr)_{y \rho}^{\nu w} \, \frac{\partial \hat{A}_\mu^z}{\partial A_\rho^w},
\end{equation}
implying the relations
\begin{equation}\label{eqn:CE}
	\frac{\partial \hat{A}_\mu^z}{\partial A_\nu^y} \, \tensor{\bigl(D^\mu(\hat{A})\bigr)}{_z^w}
	= 0
\end{equation}
and
\begin{equation}\label{eqn:311A}
	\tensor{\bigl(D_\nu(A)\bigr)}{_w^y} \, \frac{\partial\hat{A}_\mu^z}{\partial A_\nu^y}
	= 0.
\end{equation}
For an effective action $\bar{\Gamma}\bigl[\hat{A}(A)\bigr]$, depending on $A$ only via the physical gauge fields $\hat{A}$, the constraint \eqref{eqn:311A} entails the property \eqref{eqn:301A} and therefore gauge invariance. Since the constraints on $\hat{A}_\mu$ are only differential, a unique specification of $\hat{A}_\mu$ requires to fix an ``initial value'' $\bar{A}_{r,\mu}$, whose value is not relevant.

One may be tempted to employ global transversal and longitudinal fields
\begin{equation}\label{eqn:G11D}
	A^\transp_\mu
	= \tensor{P}{_\mu^\nu} A_\nu,
	\enskip
	A^L_\mu
	= A_\mu - A^\transp_\mu
	= D_\mu D^{-2} D^\nu A_\nu,
\end{equation}
which obey
\begin{equation}\label{eqn:G11E}
	D^\mu A^\transp_\mu
	= 0,
	\qquad
	D^\mu A^L_\mu
	= D^\mu A_\mu
	= \partial^\mu A_\mu.
\end{equation}
For non-abelian gauge theories they differ, however, from the physical gauge fields. Indeed, let us consider a transversal field $A_\mu$, obeying $D^\mu A_\mu = 0$. Adding an infinitesimal transversal fluctuation $f_\mu$ one finds
\begin{alignedeqn}\label{eqn:G11F}
	&D^\mu(A + f)(A_\mu + f_\mu)
	= \partial^\mu(A_\mu + f_\mu)\\
	&= D^\mu A_\mu + D^\mu f_\mu - i [f^\mu,A_\mu]
	= -i [f^\mu,A_\mu].
\end{alignedeqn}
The commutator vanishes for abelian gauge fields. For non-abelian gauge theories the physical gauge field $\hat{A}_\mu+ f_\mu$ is no longer necessarily transversal if $\hat{A}_\mu$ is transversal. (Since for a second step $f^{(2)}_\mu$ one has $D^\mu(\bar{A} + f^{(1)})f^{(2)}_\mu = 0$ the physical gauge fields do not obey $D^\mu(\bar{A})\hat{A}_\mu = 0$ either.) For non-abelian gauge theories the transversal fluctuations $f_\mu$ are related to the concept of physical fields, while general transversal gauge fields $A^\transp_\mu$ play no particular role.

Transversal abelian fields can be used for a simple construction of a subclass of physical gauge fields for non-abelian gauge theories. Indeed, fields obeying
\begin{equation}\label{eq:AS:60A} 
A_{\mu}(x)=n^{z}T_{z}B_{\mu}(x)\, ,\quad \partial^{\mu}B_{\mu}(x)=0\, ,
\end{equation}
are physical gauge fields, with reference field $\bar{A}_{r,\mu}=0$. The components $A_{\mu}^{z}(x)=B_{\mu}(x)n^{z}$ are proportional to the same $n^{z}$ for all $x$ and $\mu$, and the same holds for the infinitesimal difference $h_{\mu}^{z}(x)=(B_{\mu}^{\prime}(x)-B_{\mu}(x))\, n^{z}$. With
\begin{equation}\label{eq:AS:60B}
h_{\mu}=n^{z}T_{z}\left (B_{\mu}^{\prime}(x)-B_{\mu}(x)\right )\, ,
\end{equation}
and
\begin{align}\label{eq:AS:60C}
D^{\mu}h_{\mu}&=\partial^{\mu}h_{\mu}-i[A^{\mu},h_{\mu}]=\partial^{\mu}h_{\mu}\nonumber\\
&=n^{z}T_{z}\left (\partial^{\mu}B^{\prime}_{\mu}-\partial^{\mu}B_{\mu}\right )=0\, ,
\end{align}
one infers that $h_{\mu}=f_{\mu}$ is indeed a physical fluctuation. Furthermore, the physical gauge fields that are infinitesimally close to a reference field $\bar{A}_{r,\mu}=0$ are the transversal gauge fields. The abelian field \eqref{eq:AS:60A} can therefore indeed be constructed by a subsequent addition of physical fluctuations, starting from $\bar{A}_{r,\mu}=0$. This generalizes to linear combinations of commuting abelian fields. For example, if $T_{a}$ and $T_{b}$ commute the field $A_{\mu}=B_{\mu}^{(a)}T_{a}+B_{\mu}^{(b)}T_{b}$, $\partial^{\mu}B_{\mu}^{(a)}=0$, $\partial^{\mu}B_{\mu}^{(b)}=0$, is a physical gauge field. The difference between physical gauge fields and transversal gauge fields becomes important only for gauge fields that cannot be associated to the ones of an abelian subgroup.

For abelian fields the differential identities \eqref{eqn:275A}, \eqref{eqn:XYA}, \eqref{eqn:CC}, \eqref{eqn:CD} for the physical gauge fields become trivial. The covariant derivatives become simple partial derivatives. For a general abelian gauge field
\begin{equation}\label{eq:AS:60D} 
A_{\mu}=C_{\mu}n^{z}T_{z}
\end{equation}
the associated physical gauge field $\hat{A}(A)$ is given by the transversal part
\begin{equation}\label{eq:AS:60E} 
\hat{A}_{\mu}=\left (C_{\mu}-\partial^{-2}\partial_{\mu}\partial^{\nu}C_{\nu}\right )\, n^{z}T_{z}\, ,
\end{equation}
such that
\begin{equation}\label{eq:AS:60F}
\dfrac{\partial \hat{A}^{z}_{\mu}}{\partial A_{\nu}^{y}}=\left (\delta^{\nu}_{\mu}-\partial^{-2}\partial_{\mu}\partial^{\nu}\right )\delta_{y}^{z}=P_{\mu y}^{z\nu}.
\end{equation}
(There is no difference between $P_{\mu y}^{z\nu}(A)=P_{\mu y}^{z\nu}(\hat{A})=P_{\mu y}^{z\nu}$.) The differential identities reflect then simply the projector properties.

\subsection{Gauge-invariant effective action}

A gauge-invariant effective action depends by construction only on the gauge invariant physical gauge fields $\hat{A}_\mu$. Showing this explicitly requires, however, some care. As an example, we expand the invariant $\int_x \Tr\{F^{\mu\nu}F_{\mu\nu}\}$ for arbitrary gauge field fluctuations $h_\mu$,
\begin{align}\label{eqn:YX1}
	I
	&= \int_x \Tr\bigl\{F^{\mu\nu}(A + h)F_{\mu\nu}(A + h) - F^{\mu\nu}(A)F_{\mu\nu}(A)\bigr\}\notag\\
	&= I_1 + I_2 + \dots
\end{align}
We want to show that $I$ depends only on the difference between physical gauge fields, e.g. it vanishes if $h$ is gauge fluctuation, $h_{\mu}=D_{\mu}h$. For this purpose we employ
\begin{equation}\label{eqn:YX2}
	F_{\mu\nu}(A + h) - F_{\mu\nu}(A)
	= D_\mu h_\nu - D_\nu h_\mu - i [h_\mu,h_\nu].
\end{equation}
The term linear in $h$ appears in the field equations
\begin{equation}\label{eqn:YX3}
	I_1
	= 4 \int_x \Tr(F^{\mu\nu}{_{;\nu}}h_\mu),
\end{equation}
with
\begin{equation}\label{eqn:284A}
	F^{\mu\nu}{_{;\nu}}
	= D_\nu F^{\mu\nu}
	= \partial_\nu F^{\mu\nu} - i [A_\nu,F^{\mu\nu}].
\end{equation}
The quadratic term reads
\begin{alignedeqn}\label{eqn:YX4}
	I_2
	&= -2 \int_x \Tr h_\mu Q^{\mu\nu}h_\nu,\\
	Q^{\mu\nu}
	&= D^2\eta^{\mu\nu} - D^\mu D^\nu - 4iF^{\mu\nu}.
\end{alignedeqn}

We decompose $h_\mu$ according to
\begin{equation}\label{eqn:YX5}
	h_\mu
	= f_\mu + D_\mu \lambda,
	\qquad
	D^\mu f_\mu
	= 0.
\end{equation}
With $D_\mu D_\nu F^{\mu\nu} = 0$ one finds that $I_1$ depends only on the transversal fluctuations $f_\mu$
\begin{equation}\label{eqn:YX5A}
	I_1
	= 4 \int_x \Tr(F^{\mu\nu}{_{;\nu}}f_\mu).
\end{equation}
For the quadratic term one has
\begin{alignedeqn}\label{eqn:YX6}
	&\int_x \Tr\bigl\{(D_\mu\lambda)Q^{\mu\nu}(D_\nu\lambda)\bigr\}\\
	&= -\int_x \Tr\{\lambda D_\mu Q^{\mu\nu} D_\nu\lambda\}\\
	&= -\int_x \Tr\bigl\{\lambda\Bigl([D^\mu,D^2]D_\mu\lambda - 2iF^{\mu\nu}[D_\mu, D_\nu]\lambda\\
	&\qquad + 4i (F^{\mu\nu}{_{;\nu}})D_\mu\lambda\Bigr)\bigr\}\\
	&= - 2 i \int_x \Tr\bigl\{\lambda(F^{\mu\nu}{_{;\nu}})D_\mu\lambda\bigr\},
\end{alignedeqn}
where we use the commutator relations
\begin{alignedeqn}\label{eqn:YX7}
	[D_\mu,D_\nu]\lambda
	&= -i[F_{\mu\nu},\lambda],\\
	[D^2,D_\mu]\lambda
	&= 2i [F_{\mu\nu},D^\nu\lambda] + i [\tensor{F}{_\mu^\nu_;_\nu},\lambda].
\end{alignedeqn}
Similarly, one obtains
\begin{alignedeqn}\label{eqn:YX8}
	&\int_x \Tr\bigl\{f_\mu Q^{\mu\nu} D_\nu\lambda + (D_\mu\lambda)Q^{\mu\nu}f_\nu\bigr\}\\
	&= 2 \int_x \Tr\bigl\{f^\mu[D^2,D_\mu]\lambda + 2i[f_\mu,D_\nu\lambda] F^{\mu\nu}\bigr\}\\
	&= 2i\int_x \Tr f_\mu[F^{\mu\nu}{_{;\nu}},\lambda]\bigr\}.
\end{alignedeqn}

For $J^\mu = 0$ the field equations imply $F^{\mu\nu}{_{;\nu}} = 0$, such that the r.h.s. of eq.~\eqref{eqn:YX3}, \eqref{eqn:YX6}, \eqref{eqn:YX8} vanishes and $I_2$ depends only on the physical fluctuations $f_\mu$,
\begin{equation}\label{eqn:YX9}
	I_2
	= -2 \int_x \Tr\{f^\mu D^2 f_\mu - 4 i f_\mu F^{\mu\nu} f_\nu\}.
\end{equation}
In this case the absence of terms involving longitudinal fluctuations, that we want to establish, is seen directly. For $F^{\mu\nu}\,_{;\nu}\neq 0$ the terms \eqref{eqn:YX6} and \eqref{eqn:YX8} involve contributions to $I_{2}$ that do not vanish for $h_{\mu}=D_{\mu}h$. We will show that these terms are actually needed for the property that $I$ only involves the difference of physical gauge fields.

Indeed, the non-vanishing contributions \eqref{eqn:YX6}, \eqref{eqn:YX8} for $D_\nu F^{\mu\nu}\neq 0$ are related to the non-linear relation which connects the difference between two physical gauge fields $\hat{h}_\mu = \hat{A}_\mu^\prime - \hat{A}_\mu$ to the difference between two gauge fields or the fluctuation $h_\mu = A_\mu^\prime - A_\mu = f_\mu + D_\mu\lambda$. We write this relation in the form (for invertible $S$)
\begin{equation}\label{eqn:228Aa}
	\tensor{S}{_\mu^\nu} \, \hat{h}_\nu
	= f_\nu + E_\nu,
\end{equation}
where $E_\nu$ vanishes in linear order. Only in linear order $\hat{h}_\mu$ is independent of $\lambda$, as can be seen from the constraint \eqref{eqn:275B},
\begin{equation}\label{eqn:288B}
	\hat{h}_\mu
	= \frac{\partial \hat{A}_\mu}{\partial A_\nu} h_\nu
	= \frac{\partial \hat{A}_\mu}{\partial A_\nu} f_\nu
	= \tensor{(S^{-1})}{_\mu^\nu} f_\nu.
\end{equation}
In quadratic order $E_\nu$ is no longer purely transversal, $D^\nu E_\nu\neq 0$.

In turn, we can write
\begin{alignedeqn}\label{eqn:288C}
	I_1
	&= 4 \int_x \Tr\{D_\nu F^{\mu\nu} \tensor{S}{_\mu^\rho} \hat{h}_\rho \} + \Delta I_1,\\
	\Delta I_1
	&= -4 \int_x \Tr\{\tensor{F}{^\mu^\nu_;_\nu} E_\mu\}.
\end{alignedeqn}
For
\begin{equation}\label{eqn:288D}
	E_\mu
	= i \bigl[(f_\mu + \frac{1}{2} D_\mu\lambda),\lambda\bigr] + \dots
\end{equation}
the corresponding part in $\Delta I_1$ cancels the contributions to $I_2$ from eq.~\eqref{eqn:YX6}, \eqref{eqn:YX8}. As it should be, the difference \eqref{eqn:YX1} only depends on the physical fluctuations $\hat{h}_\mu$. (In quadratic order we can replace $f_\mu$ by $\tensor{S}{_\mu^\nu} \hat{h}_\nu$.)

In app.~\ref{app:non-linear decomposition} we present a more systematic discussion of this issue in terms of a non-linear field decomposition of $A_\mu$. This establishes in a simple form that invariants such as $\Tr F_{\mu\nu} \, F^{\mu\nu}$ only depend on the physical gauge fields $\hat{A}_\mu$.

For solutions of the vacuum field equations, $F^{\mu\nu}\,_{;\nu}=0$, the relevant operator for the quadratic fluctuations contains already implicitly a projector on the physical fluctuations. In order to see this we define the operator $\tilde{Q}^{\mu\nu}$ by
\begin{equation}\label{eqn:295A}
	\tilde{Q}^{\mu\nu}B
	= (D^2 \eta^{\mu\nu} - D^\mu D^\nu)B - 2 i [F^{\mu\nu},B],
\end{equation}
such that
\begin{equation}
	I_2
	= -2\int_x \Tr\{h_\mu\tilde{Q}^{\mu\nu}h_\nu\}.
\end{equation}
Applying the projector on the gauge fluctuations, we observe the relations
\begin{align}\label{eqn:295B}
	D^\mu D^{-2}D_\rho\tilde{Q}^{\rho\nu}B
	&= i D^\mu D^{-2}[F^{\nu\rho}{_{;\rho}},B],\\
	\label{eqn:295C}
	\tilde{Q}^{\mu\rho}D_\rho D^{-2}D^\nu B
	&= i[F^{\mu\rho}{_{;\rho}},D^{-2}D^\nu B].
\end{align}
The r.h.s. of eqs.~\eqref{eqn:295B},\eqref{eqn:295C} vanishes for $F^{\mu\nu}\,_{;\nu}$. This shows that for $F^{\mu\nu}{_{;\nu}} = 0$ the use of projectors becomes rather simple for the operator $\tilde{Q}^{\mu\nu}$ - they are not needed explicitly.

This feature is easily generalized for arbitrary gauge-invariant terms $K$. Gauge invariance implies
\begin{equation}\label{eqn:78A}
	\tensor{(D_\mu)}{_w^z} \, \frac{\partial K}{\partial A_\mu^z}
	= 0.
\end{equation}
Taking a further derivative yields
\begin{alignedeqn}\label{eqn:78B}
	\tensor{(D_\mu)}{_w^z} \, \frac{\partial^2 K}{\partial A_\mu^z \, \partial A_\nu^y}
	&= -\biggl(\frac{\partial}{\partial A_\nu^y} \tensor{(D_\mu)}{_w^z}\biggr) \frac{\partial K}{\partial A_\mu^z}\\
	&= \tensor{f}{_y_w^z} \, \frac{\partial K}{\partial A_\nu^z}
\end{alignedeqn}
For all configurations where $\partial K/\partial A_\nu^z = 0$ the second derivative of $K$ is transversal. Applying this for arbitrary linear combinations of invariants yields useful identities. In particular, if we take for $K$ a gauge invariant effective action $\bar{\Gamma}[A]$, one infers that for solutions of the field equations for $J^\mu = 0$ the second functional derivative of $\bar{\Gamma}$ is automatically transversal. No explicit projector is needed for the computation of $\Gamma_\text{P}^{(2)}$ in this case, since $\bar{\Gamma}^{(2)}$ obeys automatically the required projection properties. This constitutes an additional indication that at least for configurations of this type no spurious non-localities are introduced by the use of projectors.

\subsection{Gauge-invariant effective action and functional derivatives}

Practical computations in later parts of this paper will be performed with a simple truncation for the gauge invariant effective action. We summarize for later purposes a few properties of the relevant functional derivatives. Consider a simple form of the gauge-invariant effective action
\begin{equation}\label{eqn:YX10}
	\bar\Gamma
	= \frac{i}{2g^2} \int_x \Tr\{F^{\mu\nu} F_{\mu\nu}\},
\end{equation}
where $g$ is the gauge coupling. The first functional derivative yields the field equations for $A_\mu$, which can be inferred from eq.~\eqref{eqn:YX3}
\begin{equation}\label{eqn:YX11}
	D_\nu F^{\mu\nu}
	= -i g^2 J^\mu.
\end{equation}
A neighboring solution $A_\mu + h_\mu$ has to obey the field equations for a neighboring source $J^\mu + \delta J^\mu$. The conservation equation for $J^\mu + \delta J^\mu$ involves then the covariant derivative formed with $A_\mu + h_\mu$. In terms of covariant derivatives $D_\mu(A)$ formed with $A_\mu$ this relates two neighboring physical sources by
\begin{equation}\label{eqn:YX12}
	D_\mu(J^\mu + \delta J^\mu) - i[h_\mu,J^\mu + \delta J^\mu]
	= 0.
\end{equation}
In the linear approximation for $h_\mu$ and $\delta J^\mu$, using $D^\mu J_\mu = 0$, one finds
\begin{equation}\label{eqn:YX13}
	D_\mu \delta J^\mu
	= i [h_\mu,J^\mu].
\end{equation}

In second order in $h_\mu$ one has
\begin{alignedeqn}\label{eqn:A21-1}
	\bar\Gamma_2
	&= \frac{i}{2g^2} I_2
	= -\frac{i}{g^2} \int_x \Tr\{h_\mu \tilde{Q}^{\mu\nu}h_\nu\}\\
	&= \frac{1}{2} h^y_\mu \left(\bar \Gamma^{(2)}\right)^{\mu\nu}_{yz} h^z_\nu,
\end{alignedeqn}
with second functional derivative
\begin{equation}\label{eqn:A21-2}
	\bigl(\bar\Gamma^{(2)}\bigr)^{\mu\nu}_{yz}
	= \frac{i}{g^2}\bigl\{(-D^2 \eta^{\mu\nu} + D^\mu D^\nu)_{yz} + 2 g f_{wyz} F^{w,\mu\nu}\bigr\}.
\end{equation}
Taking into account the antisymmetry of the differential operator $\partial_\mu$ the second functional derivative $(\bar{\Gamma}^{(2)})^{\mu\nu}_{yz}(x,x^\prime)$ is symmetric in the space spanned by the indices $(\mu,y,x)$ or $(\nu,z,x^\prime)$, as it should be. From eq.~\eqref{eqn:78B} we conclude that $\Gamma^{(2)}$ is transversal for configurations $A_\mu$ with $\tensor{F}{^\mu^\nu_;_\nu}=0$, but not for arbitrary macroscopic fields $A_\mu$.

\section{Functional integral}
\label{sec:functional integral}

In this section we construct the gauge-invariant effective action from a functional integral. This proceeds largely in parallel to the usual gauge-fixing procedure in the background field formalism. Only a particular class of ``physical'' gauges can be employed, however. Furthermore, the background field is no longer an independent field. It is replaced by the macroscopic gauge field $A_\mu$, which is the argument of the gauge-invariant effective action. This leads to an implicit definition of the effective action by a functional differential equation.

\subsection{Partition function}

We split the fluctuating or microscopic fields $A_\mu^\prime$ in the functional integral into transversal and longitudinal fluctuations according to
\begin{alignedeqn}\label{eqn:G4}
	&A_\mu^\prime
	= \hat{A}_\mu + b^\prime_\mu + c^\prime_\mu,
	&&D^\mu b^\prime_\mu
	= 0,\\
	&\tilde{c}
	= D^{-2}D^\nu (A_\nu^\prime - \hat{A}_\nu),
	\qquad
	&&c^\prime_\mu
	= D_\mu \tilde{c}.
\end{alignedeqn}
Here $\hat{A}_\mu = \hat{A}_\mu(A)$ is the physical gauge field associated to the macroscopic gauge field, and covariant derivatives involve the physical field $\hat{A}_\mu$. We can write $b^\prime$ and $c^\prime$ in terms of the projector $P$,
\begin{equation}\label{eqn:G5}
	b^\prime_\mu
	= \tensor{P}{_\mu^\nu} (A_\nu^\prime - \hat{A}_\nu),
	\qquad
	c^\prime_\mu
	= \tensor{(1 - P)}{_\mu^\nu} (A_\nu^\prime - \hat{A}_\nu),
\end{equation}
where the projector $P(\hat{A})$ is formed with the physical gauge field $\hat{A}_\mu$ via the covariant derivatives $D_\mu(\hat{A})$.

We discuss here the case of a gauge fixed formulation, with partition function
\begin{align}\label{eqn:307A}
	Z
	= \int& \mathcal{D} A^\prime \, M[A^\prime,\hat{A}]\\
	&\times\exp \bigl\{-\bigl(S[A^\prime] + S_\text{gf}[A^\prime,\hat{A}] + S_L[A^\prime,L]\bigr)\bigr\}.\notag
\end{align}
Here $S[A^\prime]$ is invariant under gauge transformations of $A_\mu^\prime$ and $S_L(A^\prime,L)$ is given by eq.~\eqref{eqn:J1}. For the gauge fixing we choose a particular Landau gauge
\begin{alignedeqn}\label{eqn:307B}
	S_\text{gf}
	&= \frac{1}{\alpha} \int_x \Tr\bigl\{\bigl[(D^\mu(\hat{A})(A_\mu^\prime - \hat{A}_\mu)\bigr]^2\bigr\}\\
	&= -\frac{1}{\alpha} \int_x \Tr\big \{(A_\mu^\prime -\hat{A}_\mu) D^\mu D^\nu (A_\nu^\prime - \hat{A}_\nu)\bigr\},
\end{alignedeqn}
and take the limit $\alpha \to 0$. The covariant derivatives are again formed with the physical gauge field $\hat{A}_\mu$. Thus the gauge fixing depends on the macroscopic gauge field $A$, which will be the argument of the effective action, via the physical gauge fields $\hat{A}(A)$. (Formally, it therefore depends on the reference field $\bar{A}_r$ used for the definition of $\hat{A}(A)$.) The appearance of the macroscopic gauge field $A$ in the formulation of the partition function is an important new feature in our formalism.

We have chosen this particular gauge fixing such that $S_\text{gf}$ depends only on $c^\prime_\mu$, not on $b^\prime_\mu$,
\begin{equation}\label{eqn:307C}
	S_\text{gf}
	= \frac{1}{\alpha} \int_x \Tr\bigl\{\bigl(D^\mu(\hat{A})c^\prime_\mu\bigr)^2\bigr\}.
\end{equation}
For $\alpha \to 0$ this realizes the decoupling scenario of ref.~\cite{CWGD}, with a diverging quadratic term for $c^\prime$. Since eq.~\eqref{eqn:307C} does not involve the physical fluctuations $b_\mu^\prime$, it is a ``physical'' gauge fixing. In app.~\ref{app:non-linear decomposition} we discuss improved physical gauge fixings ($\alpha \to 0$) such as
\begin{equation}\label{eqn:76A}
	S_\text{gf}
	= \frac{1}{\alpha} \int_x \Tr\bigl\{\bigl[D^{-2}(\hat{A}) \, D^\mu(\hat{A}) \, (A_\mu^\prime - \hat{A}_\mu)\bigr]^2\bigr\},
\end{equation}
as well as an optimized physical gauge fixing where $\hat{A}$ is replaced by $\hat{A}^\prime$, the physical gauge field associated to the microscopic field $A^\prime$. For the practical discussions of this paper eq.~\eqref{eqn:76A} leads to the same result as the choice \eqref{eqn:307B}, while some conceptual issues are clearer. In the main text we concentrate on the gauge fixing \eqref{eqn:307B} because of its close connection to familiar work in Landau gauge.

The source term \eqref{eqn:J1} decomposes as
\begin{equation}\label{eqn:307D}
	S_L
	= \int_x \bigl(\tilde{L} - 2 \Tr\{J^\mu b^\prime_\mu + H^\mu c^\prime_\mu\}\bigr).
\end{equation}
It couples $b_\mu^\prime$ to the physical sources $J^\mu$ and $c^\prime_\mu$ to the ``unphysical'' sources $H^\mu$,
\begin{equation}\label{eqn:307E}
	H^\mu
	= L^\mu - J^\mu
	= D^\mu D^{-2} D_\nu L^\nu.
\end{equation}
(At this stage covariant derivatives involve $\hat{A}$. This will later be extended to arbitrary $A$.) Finally, the Faddeev-Popov determinant reads
\begin{equation}\label{eqn:307F}
	M[A^\prime,A]
	= \Det\Bigl[-\tensor{\bigl(D^\mu(\hat{A})\bigr)}{_z^w}\tensor{\bigl(D_\mu(A^\prime)\bigr)}{_w^y}\Bigr],
\end{equation}
with $D_\mu$ given by eq.~\eqref{eqn:23AA}.

So far we have not specified the definition of the macroscopic gauge field $A_\mu$. The macroscopic gauge field $A_\mu$ and the sources $L^\mu$ should be in some fixed relation, $A_\mu(L^\mu)$. For example, we could take
\begin{equation}\label{eqn:YA1}
	A_\mu^z(x)
	= \frac{\partial \ln Z[L,A]}{\partial L_z^\mu(x)}\Bigr|_A,
\end{equation}
which identifies the macroscopic field $A_\mu$ with the expectation value of the microscopic field $A_\mu^\prime$ for the source $L$,
\begin{equation}\label{eqn:YA2}
	A_\mu
	= \langle A_\mu^\prime\rangle_L.
\end{equation}
We will adopt a different choice for the relation between $A$ and $L$, which will be specified below. The relation \eqref{eqn:YA1} turns the definition of $Z$ in eq.~\eqref{eqn:307A} into an implicit integro-differential equation, since $A_\mu(L)$ involves partial derivatives of $Z$. This will generalize to other choices of the relation between $A$ and $L$. We will not have to solve this type of equation explicitly.

Our setting resembles in many aspects the construction in the background field formalism in ref.~\cite{RW}, but there are also important differences. The major difference concerns the absence of an independent background field. For the gauge fixing the background field is replaced by the physical gauge field $\hat{A}(A)$ for the covariant derivative and the expansion point entering the definition of $c_\mu^\prime$. A different choice of source is the second important difference to the formulation in ref.~\cite{RW}. As a third difference to the construction in ref.~\cite{RW} the gauge fixing is not arbitrary but restricted to physical gauge fixing terms.

\subsection{Effective action}

We define the effective action $\tilde\Gamma[A]$ by the implicit expression
\begin{align}\label{eqn:346A}
	&\exp\bigl(-\tilde\Gamma[A]\bigr)
	= \int \mathcal{D} A^\prime \, M[A^\prime,A]\\
	&\hphantom{{}=}\times\exp \biggl\{-\bigl(S[A^\prime] + S_\text{gf}[A^\prime,A]\bigr) + \int_xL^\mu_z(A^{\prime z}_\mu - A^z_\mu)\biggr\}\notag\\
	&\hphantom{\exp\bigl(-\tilde\Gamma[A]\bigr)}= -\bigl(\ln Z + S_L[L,A]\bigr),\notag
\end{align}
where $L(A)$ denotes the source associated to $A$. Our choice of the relation between $A$ and $L$ is given by
\begin{equation}\label{eqn:96A}
	L_z^\mu
	= \frac{\partial \tilde\Gamma}{\partial A_\mu^z}.
\end{equation}
This results in the central functional differential equation
\begin{alignedeqn}\label{eqn:96B}
	&\exp\bigl(-\tilde\Gamma[A]\bigr)
	= \int \mathcal{D} A^\prime\\
	&\times \exp\biggl\{-\tilde{S}[A^\prime,A] + \int_x \frac{\partial \tilde\Gamma}{\partial A_\mu^z} \, \bigl(A_\mu^{\prime z} - A_\mu^z\bigr)\biggr\},
\end{alignedeqn}
with
\begin{equation}\label{eqn:96C}
	\tilde{S}[A^\prime,A]
	= S[A^\prime] + S_\text{gf}[A^\prime,A] - \ln M[A^\prime,A].
\end{equation}
Since eq.~\eqref{eqn:96B} is a differential relation, one needs, in principle, the specification of boundary conditions for a unique definition of $\tilde\Gamma[A]$ and $L(A)$. This may be given by $\tilde\Gamma[\bar{A}_r] = 0$ or $\tilde\Gamma[A = 0] = 0$.

We write $\tilde\Gamma[A]$ in the form
\begin{equation}\label{eqn:347A}
	\tilde\Gamma[A]
	= \bar{\Gamma}[\hat{A}] + \Gamma_\text{gf} [\hat{A},\hat{c}] + \Delta\Gamma [\hat{A},\hat{c}],
\end{equation}
where $\Delta\Gamma$ vanishes for $\hat{c} = 0$. Here we employ the decomposition \eqref{eqn:274A} of $A_{\mu}$ into physical fields $\hat{A}_\mu$ and gauge degrees of freedom $\hat{c}_\mu$. The gauge-fixing term
\begin{equation}\label{eqn:307G}
	\Gamma_\text{gf}
	= \frac{1}{\alpha} \int_x \Tr\bigl\{\bigl(D^\mu(\hat{A})\hat{c}_\mu\bigr)^2\bigr\}
\end{equation}
will turn out to be the only term which diverges for $\alpha \to 0$.

\subsection{Field equation in gauge sector}

We next want to show that in the limit $\alpha \to 0$ the partial solution of the field equations for the gauge degrees of freedom implies $\hat{c}_\mu = 0$. Inserting this solution into the effective action $\tilde\Gamma[A]$ results in the gauge-invariant expression $\tilde\Gamma[\hat{A},\hat{c} = 0] = \bar\Gamma[\hat{A}]$. This will be the basis for the definition of the gauge-invariant effective action.

For this purpose we write eq.~\eqref{eqn:346A} in the form
\begin{equation}\label{eqn:346C}
	\tilde\Gamma[A]
	= \frac{1}{\alpha} \int_x \Tr\bigl\{\bigl(D^\mu(\hat{A})\hat{c}_\mu\bigr)^2\bigr\} + F[A],
\end{equation}
with
\begin{alignedeqn}\label{eqn:346D}
	F[A]
	&= -\ln\int_x \mathcal{D} A^\prime M[A^\prime,A] \, B[c^\prime,A]\\
	&\hphantom{{}=}\times \exp \bigl\{-S[A^\prime] + 2 \int_x \Tr(J^\mu b^\prime_\mu - L^\mu\hat{A}_\mu)\bigr\},
\end{alignedeqn}
and
\begin{alignedeqn}\label{eqn:346E}
	B[c^\prime,A]
	&= \exp\Biggl[\int_x \Tr\biggl\{\left(2H^\mu + \frac{2}{\alpha} D^\mu D^\nu\hat{c}_\nu\right)\\
	&\hphantom{{}=}\times (c^\prime_\mu - \hat{c}_\mu)
	- \frac{1}{\alpha} \bigl(D^\mu(c^\prime_\mu - \hat{c}_\mu)\bigr)^2\biggr\}\Biggr].
\end{alignedeqn}
Here $\hat{A}_\mu$ is related to $A_\mu$ and $\hat{c}_\mu$ by $\hat{A}_\mu = A_\mu - \hat{c}_\mu$ and we do not use at this point the properties of physical gauge fields. The precise choice of $\hat{c}_\mu$ is given below. For establishing the leading term \eqref{eqn:307G} it is sufficient to show that $F$ remains finite for $\alpha \to 0$.

We next proceed to a saddle point expansion and expand $B$ around its extremum, which occurs for $c^\prime = c_0$,
\begin{equation}\label{eqn:346F}
	D^\mu D^\nu c_{0,\nu}
	= -\alpha H^\mu.
\end{equation}
Identifying
\begin{equation}\label{eqn:346G}
	\hat{c}_\mu
	= c_{0,\mu}
\end{equation}
the factor $B$ becomes unity, such that the leading order saddle point approximation does not produce in $\tilde\Gamma$ any additional terms diverging $\sim \alpha^{-1}$. Insertion of eq.~\eqref{eqn:346F}, \eqref{eqn:346G} yields
\begin{equation}\label{eqn:346H}
	B[c^\prime,A]
	= \exp\left(-\frac{1}{\alpha} \int_x \Tr\bigl[D^\mu(c^\prime_\mu - \hat{c}_\mu)\bigr]^2\right).
\end{equation}
For $\alpha \to 0$ one finds for any finite source $H^\mu$ the simple solution $\hat{c}_\mu = c_{0,\mu} = 0$. The lowest-order saddle point approximation for $F[A]$ inserts $\hat{c} = 0$ in eq.~\eqref{eqn:346D}, resulting in the replacement $A \to \hat{A}$. This implies in eq.~\eqref{eqn:347A} $\Delta\Gamma = 0$, with $\bar\Gamma[\hat{A}] = F[A=\hat{A}]$. For $\alpha \to 0$ and infinitesimal $\hat{c}_\mu$ the longitudinal character of $\hat{c}_\mu$ according to eq.~\eqref{eqn:346G} is sufficient to show that $\hat{c}_\mu$ is indeed a gauge fluctuation. Thus $\hat{A}_\mu$ is a physical gauge field.

Higher-order terms in the saddle point approximation do not produce terms diverging $\sim \alpha^{-1}$. This may be seen in a somewhat sketchy way by decomposing the functional measure
\begin{equation}\label{eqn:346I}
	\int \mathcal{D} A^\prime
	= \int \mathcal{D}\tilde{b} \, \mathcal{D}\tilde{c} \, N(A),
\end{equation}
where $\tilde{b}$ and $\tilde{c}$ are unconstrained fields formed from $b^\prime$ and $c^\prime$, respectively \cite{CWGD}, and $N(A)$ is a normalization factor. The factor $B$ in eq.~\eqref{eqn:346D}, \eqref{eqn:346H} becomes
\begin{equation}
	B
	= \exp\biggl\{-\int_x \frac{1}{2 \alpha} (\tilde{c} - \tilde{c}_0) \, D^4(A) \, (\tilde{c} - \tilde{c}_0)\biggr\}.
\end{equation}
Making a variable change $\tilde{d} = (\tilde{c} - \tilde{c}_0)/\sqrt{\alpha}$ absorbs the $\alpha$-dependence in $\int \mathcal{D} A^\prime \, B$ into a field-independent part of the Jacobian that can be neglected. All other dependence on $\tilde{c}$ in $M[A^\prime,A]$, $S[A^\prime]$, etc. appears now with appropriate factors of $\sqrt{\alpha}$. This can be neglected for $\alpha \to 0$, where the saddle point approximation in the $c^\prime$-sector becomes exact. This concludes the argument that $\Delta \Gamma$ in eq.~\eqref{eqn:347A} remains finite for $\alpha \to 0$, and therefore $\hat{c}_\mu = 0$ for all solutions of the field equations with finite sources.

\subsection{Gauge-invariant effective action}

The gauge-invariant effective action is obtained by extending $\bar\Gamma[\hat{A}]$ to $\bar\Gamma[A]$, associating to every macroscopic field $A_\mu$ the corresponding gauge-invariant field $\hat{A}_\mu$,
\begin{alignedeqn}\label{eqn:307H}
	\bar{\Gamma}[A]
	= \bar{\Gamma}\bigl[\hat{A}(A)\bigr],
	\qquad
	\bar{\Gamma}[\hat{A}]
	= \tilde\Gamma[A]\bigr|_{\hat{c}=0}.
\end{alignedeqn}
Gauge invariance of $\bar\Gamma[A]$ reflects directly that $\bar\Gamma$ depends on $A$ only through its dependence on the physical gauge field $\hat{A}(A)$ \cite{CWGD}. The construction \eqref{eqn:307H} eliminates the gauge-fixing term as well as all other terms involving $\hat{c}$.

The gauge-invariant effective action $\bar\Gamma[\hat{A}]$ can be obtained from the implicit definition \eqref{eqn:96B} by restricting the argument to $A = \hat{A}$. We can subsequently extend the argument of $\bar\Gamma$ to arbitrary $A$ according to eq.~\eqref{eqn:307H}. The derivative $\partial \tilde\Gamma/\partial A_\mu$ is replaced by $\partial \bar\Gamma/\partial A_\mu$, which yields the conserved source $J^\mu$. Here $\partial \bar{\Gamma}/\partial A$ has to be evaluated at $\hat{A}(A)$. The gauge-invariant effective action $\bar\Gamma[A]$ can then be defined as the gauge-invariant solution of eq.~\eqref{eqn:96B}. In summary, the gauge invariant effective action $\bar{\Gamma}[A]$ is defined by the implicit relation
\begin{alignedeqn}\label{eqn:113A}
	&\exp\bigl(-\bar\Gamma[A]\bigr)
	= \int \mathcal{D} A^\prime \, \\
	&\times \exp\biggl\{-\tilde{S}[A^{\prime},\hat{A}]+ \int_x \frac{\partial \bar\Gamma}{\partial A_\mu^z} \, \bigl(A_\mu^{\prime z} - \hat{A}_\mu^z\bigr)\biggr\},
\end{alignedeqn}
with $\tilde{S}[A^{\prime},\hat{A}]$ given by eq.~\eqref{eqn:96C} and $\hat{A}=\hat{A}(A)$.

Nevertheless, our formulation contains a gauge fixing, necessary to render the functional integral well defined. Due to the restriction on physical gauges the usual issue of the dependence of the effective action on the choice of the gauge fixing is largely absent. A small residual dependence could result from the precise selection of the physical gauge fixing. For the example of the improved physical gauge fixing \eqref{eqn:76A} the gauge invariant effective action is determined by
\begin{alignedeqn}\label{eqn:101A}
	&\exp\bigl(-\bar\Gamma[A]\bigr)
	= \int \mathcal{D} A^\prime \, \delta(\tilde{c}) \, \tilde{M}[A^\prime,\hat{A}]\\
	&\times \exp\biggl\{-S[A^\prime] + \int_x \frac{\partial \bar\Gamma}{\partial A_\mu^z} \, \bigl(A_\mu^{\prime z} - \hat{A}_\mu^z\bigr)\biggr\},
\end{alignedeqn}
with $\tilde{M}[A^\prime,\hat{A}] = \tilde{M}[A^\prime,\hat{A}(A)]$ given by eq.~\eqref{eqn:3A}. We note that the Faddeev-Popov determinant $\tilde{M}$ equals unity for $A^\prime = \hat{A}$. With
\begin{alignedeqn}\label{eqn:101B}
	&\delta(\tilde{c})
	= \prod_z \delta(G^z),\\
	&G^z
	= \bigl[D^{-2}(\hat{A}) \, D^\mu(\hat{A}) (A_\mu^\prime - \hat{A}_\mu)\bigr]^z,
\end{alignedeqn}
this yields eq.~\eqref{eqn:AA1}.

The procedure for computing the gauge invariant effective action first selects a suitable set of physical fields $\hat{A}$ and evaluates eq.~\eqref{eqn:113A} for $\bar{\Gamma}[\hat{A}]$. This is subsequently extended to $\bar{\Gamma}[A]$. In practice, $\bar{\Gamma}[A]$ will only be evaluated in a given truncation, making some gauge invariant ansatz with unknown functions. The set of physical configurations $\hat{A}$ has to be chosen large enough such that the free functions are determined once $\bar{\Gamma}[\hat{A}]$ is known for this set. Generalized perturbative expansions correspond to iterative solutions of the functional differential equation \eqref{eqn:113A}. One starts with a lowest order guess of the form of $\Gamma[A]$, for example taking the form of the classical action. This lowest order can be used for $\partial\Gamma /\partial A$ on the r.h.s. of eq.~\eqref{eqn:113A}. Then eq.~\eqref{eqn:113A} becomes a functional integral that determines the first order form of $\Gamma[A]$. Employing this first order form for $\partial\Gamma /\partial A$ one proceeds iteratively. We will below describe a computation of $\bar{\Gamma}[A]$ by a functional flow equation and proceed to a detailed discussion of its form for high covariant momenta of the gauge fields.

We finally recall that $\Delta \Gamma$ in eq.~\eqref{eqn:347A} remains finite for $\alpha \to 0$. For $\alpha \to 0$ it has no influence on the field equation for $\hat{c}$ and vanishes once the partial solution $\hat{c} = 0$ is inserted. For the derivation of the field equations, or the evaluation of $\tilde{\Gamma}$ for a solution of the field equations, we can omit the term $\Delta\Gamma$. The issue is more subtle for higher functional derivatives. For example, the second functional derivative of $\Delta\Gamma$ may not vanish of $\alpha\to 0$, $\hat{c}\to 0$, since a term linear in $\hat{c}$ could give rise to mixed derivatives. Such terms will often be eliminated by projections on physical fluctuations. For many practical purposes we can simply omit $\Delta\Gamma$. Then the gauge-fixed effective action decomposes for $\alpha \to 0$ effectively into a gauge-invariant part $\bar\Gamma[\hat{A}(A)]$ and a simple gauge-fixing term
\begin{equation}\label{eqn:108A}
	\tilde\Gamma[A]
	= \bar\Gamma[A] + \frac{1}{\alpha} \int_x \Tr\bigl\{\bigl[D^\mu(\hat{A}) \, (A_\mu - \hat{A}_\mu)\bigr]^2\bigr\}.
\end{equation}

In eq.~\eqref{eqn:113A} or \eqref{eqn:101A} we may replace $\hat{A}(A)$ by $A$ in the terms that do not diverge for $\alpha\to 0$, e.g. in $\tilde{M}$ or $(\partial\bar{\Gamma}/\partial A)(A^{\prime}-\hat{A})$. The reason is that the difference between these expressions evaluated at $\hat{A}(A)$ or at $A$ is at least linear in $\hat{c}$. Since these differences do not diverge for $\alpha\to 0$ they only modify $\Delta\Gamma$. For all issues for which $\Delta\Gamma$ can be neglected the difference therefore does not matter. For all terms in the functional integral except for the physical gauge fixing term we can then set $\hat{c}=0$.

\subsection{Field equation and expectation value of microscopic gauge field}

The field equations relate the first functional derivative of $\bar\Gamma[A]$ to the conserved source
\begin{equation}\label{eqn:108B}
	\frac{\partial \bar\Gamma}{\partial A_\mu}
	= J^\mu.
\end{equation}
On the other hand, the derivative of the gauge-fixing term in $\tilde\Gamma$ determines the field equation in the gauge sector. It enforces $A_\mu = \hat{A}_\mu$ for arbitrary finite unphysical sources $H^\mu$. Insertion of this solution eliminates the gauge-fixing term. Eq.~\eqref{eqn:108B} is the definition of the field equation for classical field theories in a quantum context. For QED it constitutes the modification of Maxwell's equations by the Euler-Heisenberg term, or more general effects from quantum fluctuations of charged particles.

In the standard construction of the effective action by a Legendre transformation of the generating functional $W[L]$ for the connected Green's functions the argument $A$ of $\tilde{\Gamma}[A]$ is directly given by the expectation value of the microscopic field, $A=\langle A^{\prime}\rangle$, c.f. eq.~\eqref{eqn:YA2}. In general, this does not hold for our implicit definition and the relation \eqref{eqn:96A} between sources and macroscopic gauge fields. One may be interested in the expectation value $\langle A^{\prime}\rangle$, even though this does not play an important role in practice. In app.~\ref{app:expectation values}, we discuss how expectation values are computed for effective actions defined by an implicit relation of the type \eqref{eqn:96B}. We find the relation
\begin{equation}\label{eqn:114A}
	\bigl(\tilde\Gamma^{(2)}\bigr)_{zy}^{\mu\nu} \, \langle A_\nu^{\prime y} - A_\nu^y\rangle
	= V_z^\mu,
\end{equation}
with
\begin{equation}\label{eqn:114B}
	V_z^\mu
	= \biggl\langle\frac{\partial \bar{S}}{\partial A_\mu^z}\biggr\rangle,
	\qquad
	\bar{S}
	= S_\text{gf} - \ln M,
\end{equation}
reflecting the dependence of the gauge-fixing term and Faddeev-Popov determinant on the macroscopic field $A$ at fixed microscopic field $A^\prime$. With $A^\prime - \hat{A} = b^\prime + c^\prime - \hat{c}$ we can use the previous result $\langle c^\prime\rangle = \hat{c} = 0$, such that eq.~\eqref{eqn:114A} fixes $\langle b^\prime\rangle$ by
\begin{equation}\label{eqn:114C}
	\bigl(\bar\Gamma_P^{(2)}\bigr)_{zy}^{\mu\nu} \, \langle b_\nu^{\prime y}\rangle
	= V_z^\mu,
\end{equation}
or
\begin{equation}\label{eqn:114D}
	\langle b_\mu^{\prime z}\rangle
	= \bigl(G_P\bigr)_{zy}^{\mu\nu} \, V_y^\nu.
\end{equation}
Due to the projection properties of $G_P$ any longitudinal part of $V_z^\mu$ does not contribute in eq.~\eqref{eqn:114D}.

For an ``optimal physical gauge fixing'' the ``source correction'' $V_z^\mu$ vanishes or is purely longitudinal. In this case one has $\langle b^\prime\rangle = 0$, such that the expectation value for the microscopic field $\langle A_\mu^\prime\rangle$ for the source given by eq.~\eqref{eqn:96A} equals the physical macroscopic gauge field $\hat{A}_\mu$,
\begin{equation}\label{eqn:103A}
	\langle A_\mu^\prime\rangle
	= \hat{A}_\mu.
\end{equation}
For any given macroscopic field $A_\mu$ one can find a reference field $\bar{A}_{r,\mu}$ such that $\hat{A}_\mu = A_\mu$. (This is achieved by gauge transformations such that $A_\mu$ is the physical field representing the gauge orbit.) For optimal physical gauge fixing we can associate $\langle A^\prime\rangle$ with the macroscopic field $A$ modulo gauge transformations. For any given conserved source $J$ the expectation value $\langle A^\prime\rangle$ is uniquely defined only if a reference field $\bar{A}_r$ is specified.

Landau gauge fixing is not optimal in this sense. The properties of $V_z^\mu$ for Landau gauge fixing are discussed in app.~\ref{app:expectation values}. For $\langle b^\prime\rangle \neq 0$ the macroscopic gauge field that solves the field equations for a given source $J$ does not equal the expectation value of the microscopic gauge field. We advocate that it is the macroscopic gauge field that matters for practical purposes, and its precise relation to $\langle A^\prime\rangle$ is of secondary importance.

At this point we note the possibility to add to $\bar{S}$ in eq.~\eqref{eqn:114B} a gauge-invariant ``correction term'' $C[A]$ that only depends on the macroscopic gauge field $A$. Being independent of $A^\prime$ the term $C[A]$ does not change the property that the physical gauge fixing acts only on the gauge fluctuations, and that the partial solution of the field equation in the gauge sector amounts to $\langle c^\prime\rangle = \hat{c} = 0$. The correction term $C[A]$ modifies $V_z^\mu$ in eq.~\eqref{eqn:114B} and therefore the relation between $\langle A^\prime\rangle$ and $A$. One may try to find a suitable functional $C[A]$ such that $\langle A^\prime\rangle = A$ also holds for Landau gauge. The correction term adds directly to $\bar\Gamma[A]$ by replacing in eq.~\eqref{eqn:101A} the factor $\delta(\tilde{c}) \, \tilde{M}$ by $\delta(\tilde{c}) \, \tilde{M} \, \exp\bigl\{-C\bigr\}$. The field equation \eqref{eqn:108B} is modified by the new definition of $\partial \bar\Gamma/\partial A$.

An extended notion of an optimal physical gauge fixing is realized if eq.~\eqref{eqn:103A} is replaced by
\begin{equation}\label{eqn:122AA}
	\langle A_\mu^\prime\rangle
	= \tilde{A}_\mu,
\end{equation}
with $\tilde{A}$ related to $\hat{A}$ by a gauge transformation in a unique way. We discuss in app.~\ref{app:non-linear decomposition} an optimal gauge fixing of this type.

\subsection{Propagator and physical correlation function}

The propagator for physical gauge field fluctuations is determined by the second functional derivative $\bar{\Gamma}^{(2)}$ of the gauge invariant effective action. More precisely, it is given by the inverse of $\bar{\Gamma}^{(2)}$ in the projected space of physical fluctuations. We may first employ eq.~\eqref{eqn:108A} for computing propagators for gauge fields and show subsequently that $\Delta\Gamma$ can indeed be neglected. The second functional derivative of $\tilde\Gamma$, evaluated at $\hat{c} = 0$, can be projected into different subsectors. One finds, neglecting relative corrections $\sim \alpha$,
\begin{alignedeqn}\label{eqn:347B}
	(1 - P^\transp) \tilde\Gamma^{(2)} (1 - P)
	&= (1 - P^\transp) \Gamma^{(2)}_\text{gf} (1 - P),\\
	(\Gamma^{(2)}_\text{gf})^{\mu\nu}_{zy}
	&= -\frac{1}{\alpha} (D^\mu D^\nu)_{zy},
\end{alignedeqn}
and
\begin{equation}\label{eqn:347C}
	P^\transp \tilde\Gamma^{(2)} P
	= P^\transp \bar\Gamma^{(2)} P
	= \bar{\Gamma}^{(2)}_P.
\end{equation}
For the projection onto the physical fluctuations we can use the gauge-invariant effective action $\bar\Gamma$, neglecting the gauge-fixing term. For the effective action \eqref{eqn:YX10} the inverse physical propagator $\bar{\Gamma}^{(2)}_P$ is given by eq.~\eqref{eqn:A21-2}.

The propagator defined by the inverse of $\tilde{\Gamma}^{(2)}$ describes the response of the solutions of the field equation \eqref{eqn:96A} to a small change of the sources. Let $A_\mu$ be the solution of the field equation \eqref{eqn:96A} for source $L^\mu$, and $A_\mu + \delta A_\mu$ the solution for source $L^\mu + \delta L^\mu$. Expanding eq.~\eqref{eqn:96A} for small $\delta A_\mu$ and $\delta L^\mu$ yields
\begin{equation}\label{eqn:CF1}
	\delta L_\mu^z
	= \frac{\partial^2 \tilde\Gamma}{\partial A_\mu^z \, \partial A_\nu^y} \, \delta A_\nu^y,
\end{equation}
or
\begin{equation}\label{eqn:CF2}
	\delta A
	= \bigl(\tilde\Gamma^{(2)}\bigr)^{-1} \, \delta L.
\end{equation}
For $\alpha \to 0$ the propagator vanishes except for the piece corresponding to the physical fluctuations. Therefore only the variation $\delta J$ of the conserved physical source contributes, and $\delta A$ is a physical fluctuation, $\delta A = \delta \hat{A}$, with
\begin{equation}\label{eqn:CF3}
	\delta A
	= G_P \, \delta J.
\end{equation}
Here the propagator $G_{P}$ for the physical fluctuations is given by inversion on the projected subspace,
\begin{equation}\label{eqn:CF4}
	\bar\Gamma_P^{(2)} \, G_P
	= P^\transp,
\end{equation}
with
\begin{equation}\label{eq:AS:129A} 
\bar{\Gamma}_{P}^{(2)}=P^{T}\bar{\Gamma}^{(2)}P.
\end{equation}

With eq.~\eqref{eqn:108A}, the second functional derivative $\tilde{\Gamma}^{(2)}$ is block diagonal according to eqs.~\eqref{eqn:347B}, \eqref{eqn:347C}. Adding the term $\Delta\Gamma$ may induce off-diagonal terms that remain, however, finite in the limit $\alpha\to 0$. The contribution of such off-diagonal terms to the propagator vanishes for $\alpha\to 0$. They can therefore be omitted. 

For theories without local gauge symmetry the propagator can be identified with the connected two-point correlation function. For local gauge theories in the background field formalism the identification of the two-point correlation function and the propagator holds as well, with $\Gamma_\text{bg}^{(2)}$ corresponding to the second functional derivative with respect to $A$ at fixed $\bar{A}$. In general, $\Gamma_\text{bg}^{(2)}$ cannot be expressed by the second derivative of $\bar\Gamma_\text{bg}[A] = \Gamma_\text{bg}[A,\bar{A} = A]$, since the latter also involves derivatives of $\Gamma_\text{bg}[A,\bar{A}]$ with respect to $\bar{A}$ at fixed $A$.

For an implicit definition of the gauge invariant effective action \eqref{eqn:113A} the identification of the propagator $G_P$ defined by eq.~\eqref{eqn:CF4}, with the correlation function for physical fluctuations $\tilde{G}_P$, as defined by ($b = \langle b^\prime\rangle$)
\begin{equation}\label{eqn:CF5}
	(\tilde{G}_P)_{\mu\nu}^{zy}(x,y)
	= \bigl\langle[b_\mu^{\prime z}(x) - b_\mu^z(x)] [b_\nu^{\prime y}(x) - b_\nu^y(x)]\bigr\rangle,
\end{equation}
is not obvious a priori. In app.~\ref{app:correlation function} we discuss conditions for the identity
\begin{equation}\label{eqn:CF6}
	\tilde{G}_P
	= G_P.
\end{equation}
Instead of optimizing the precise formulation of the gauge invariant effective action in order to achieve eq.~\eqref{eqn:103A}, one may optimize in order to realize eq.~\eqref{eqn:CF6} \cite{CWGIF}. The identity \eqref{eqn:CF6} relies then on an optimal physical gauge fixing and would not hold for general gauge fixing. For quantum gravity, an identity of this type is the basis for the computation of primordial cosmic fluctuations from the quantum effective action \cite{CWPF1,CWPF2}.

\subsection{Independence of reference field}

The definition of the physical gauge field $\hat{A}$ depends on the choice of a reference field $\bar{A}_r$ from which $\hat{A}$ is constructed by adding physical fluctuations. (We add here the subscript $r$ in order to avoid confusion with the background field in the background formalism.) Through the definition of $\hat{A}$ the gauge-fixing term $S_\text{gf}$ in $\tilde{S}$ in eq.~\eqref{eqn:96C} implicitly depends on $\bar{A}_r$. Therefore the functional differential equation \eqref{eqn:96B} depends formally on the reference field. One may ask to what extent the solution $\tilde\Gamma[A]$ depends on $\bar{A}_r$.

Different choices of $\bar{A}_r$ correspond to different physical gauge-fixing conditions. While the expression \eqref{eqn:307G} of the gauge-fixing term $\Gamma_\text{gf}$ in terms of $\hat{c}$ is the same for all choices of $\bar{A}_r$, the functional relation $\hat{c}(A)$ depends on $\bar{A}_r$. The partial solution of the field equation is independent of $\bar{A}_r$, however. All arguments following eq.~\eqref{eqn:307G} are the same for any choice of the reference field. We conclude that the gauge-invariant effective action is independent of the choice of the reference field $\bar{A}_r$. The reference field is only needed if we want to define a unique expectation value $\langle A^\prime\rangle$ according to eq.~\eqref{eqn:103A}.

\subsection{Comparison with background formalism}

It is instructive to compare our implicit definition \eqref{eqn:96B} of the effective action with the background field formalism. In the background field formalism the effective action $\tilde\Gamma_\text{bg}[A,\bar{A}]$ obeys a relation similar to eq.~\eqref{eqn:96B}, but now involving the background field $\bar{A}$ instead of the macroscopic physical field $\hat{A}(A)$. In eq.~\eqref{eqn:96C} one replaces $M[A^\prime,\hat{A}(A)]$ by $M[A^\prime,\bar{A}]$, while $S_\text{gf}[A^\prime,\hat{A}(A)]$ is replaced by $S_\text{gf}[A^\prime,\bar{A}]$, with $\bar{A}$ the fixed background field. For the source term, $\partial \tilde\Gamma/\partial A$ is replaced by the partial derivative at fixed $\bar{A}$, $\smash{\partial \tilde\Gamma/\partial A\bigr|_{\bar{A}}}$.

The effective action $\tilde\Gamma_\text{bg}[A,\bar{A}]$ depends on both $A$ and $\bar{A}$. A gauge-invariant part
\begin{equation}\label{eqn:103B}
	\bar\Gamma_\text{bg}[A] = \tilde\Gamma_\text{bg}[A,\bar{A} = A]
\end{equation}
can be defined by identifying the background field with the macroscopic field. We observe that $\bar\Gamma_\text{bg}[A]$ and $\bar\Gamma[A]$ obey almost the same relation \eqref{eqn:96B}. Indeed, for $\hat{c} = 0$ one has $\hat{A} = A$, such that $\tilde{S}[A^\prime,A]$ in eq.~\eqref{eqn:96B} is identical for $\bar\Gamma_\text{bg}[A]$ and $\bar\Gamma[A]$. The only difference is the source term, which is given for the background effective action by
\begin{equation}\label{eqn:103C}
	\frac{\partial \tilde\Gamma_\text{bg}[A,\bar{A}]}{\partial A}\biggr|_{\bar{A}}
	= \frac{\partial \bar\Gamma_\text{bg}}{\partial A} - \frac{\partial \tilde\Gamma_\text{bg}[A,\bar{A}]}{\partial \bar{A}},
\end{equation}
instead of $\partial \bar\Gamma/\partial A$ for our definition of the gauge-invariant effective action. Possible differences between our definition of the gauge-invariant effective action $\bar\Gamma[A]$ and the gauge-invariant background effective action $\bar\Gamma_\text{bg}[A]$ are rooted in the different relation between the macroscopic gauge field $A$ and the sources.

Since the unphysical sources $H$ have no influence for $\alpha \to 0$, one has
\begin{equation}\label{eqn:96D}
	\bar{\Gamma}[A]
	= \bar\Gamma_\text{bg}[A]
	= \tilde\Gamma_\text{bg}[A,\bar{A} = A]
\end{equation}
for all configurations $A$ obeying in the background field formalism the relation
\begin{equation}\label{eqn:96F}
	\tensor{P}{_\mu^\nu} \, \frac{\partial \tilde\Gamma_\text{bg}}{\partial \bar{A}_\nu^z}\Bigr|_{A = \bar{A}}
	= 0.
\end{equation}
Only for these configurations is the physical source identical for both formulations,
\begin{equation}\label{eqn:96E}
	J_z^\mu
	= \frac{\partial \bar\Gamma}{\partial A_\mu^z}
	= \frac{\partial \tilde\Gamma_\text{bg}}{\partial A_\mu^z}\Bigr|_{\bar{A} = A},
\end{equation}
and the correction to the field equations discussed in ref.~\cite{RW} vanishes. For configurations obeying eq.~\eqref{eqn:96E} both the gauge-invariant effective action and the ``classical'' field equations can be computed equivalently in our manifestly gauge-invariant approach and in the background formalism.

For general $A$ the relation \eqref{eqn:96F} needs not to hold. (This concerns, in particular, the flowing action discussed in the next section.) In this case the sources corresponding to a given macroscopic field differ between the background field formalism and our definition of the gauge-invariant effective action. In the background field formalism we may define
\begin{alignedeqn}\label{eqn:106A}
	&J_\text{bg}^\mu[A,\bar{A}]
	= \tensor{P}{_\mu^\nu}(A) \, \frac{\partial \tilde\Gamma}{\partial A_\nu}\Bigr|_{\bar{A}},\\
	&J_\text{bg}^\mu[A]
	= J_\text{bg}^\mu[A,\bar{A} = A].
\end{alignedeqn}
If eq.~\eqref{eqn:96F} does not hold for all $A$, the functional relation $J_\text{bg}^\mu[A]$ differs from $J[A]$ as defined by eq.~\eqref{eqn:108B}.

In the background formalism $A$ corresponds to the expectation value $\langle A^\prime\rangle$ for the source $J_\text{bg}[A]$. Suppose that one can find an optimal gauge fixing for our formulation of the gauge-invariant action, such that $A$ equals the expectation value $\langle A^\prime\rangle$ for the source $J$. For both formulations $A$ corresponds then to the expectation value $\langle A^\prime\rangle$. These are, however, expectation values for different sources. For a given $A$ the expectation value $\langle A^\prime\rangle_\text{bg}$ differs from $\langle A^\prime\rangle$ as evaluated in our formalism for the source $J$. This extends to correlation functions. For a given $A$ the correlation function in the background field formalism differs from the physical correlation function proposed in this paper.

One may define in the background field formalism a modified conserved source
\begin{equation}
	\bar{J}_\text{bg}^\mu
	= \frac{\partial \bar\Gamma_\text{bg}}{\partial A_\mu},
\end{equation}
which differs from $J_\text{bg}$ by the omission of the second term in eq.~\eqref{eqn:103C}. Then the macroscopic field $A_\text{bg}$ corresponding to $\bar{J}_\text{bg}$ differs from $A$ which corresponds to $J_\text{bg}$. If we choose $\bar{J}_\text{bg} = J$ the expectation value $\langle A^\prime\rangle_\text{bg}[\bar{J}_\text{bg}]$ equals $\langle A^\prime\rangle[J]$. This extends to all correlation functions. The correlation functions depend on the sources and the gauge fixing, but should not depend on the formalism used to compute them. Since the gauge fixing is the same, the correlation functions should coincide if the sources are the same.

In the background field formalism one can compute, in principle, for any $A$ the associated source $\bar{J}_\text{bg}$ and therefore $A_\text{bg}$, thus establishing a functional relation $A_\text{bg}[A]$. Then the physical correlation functions computed in our formalism for a macroscopic field $A$ should coincide with the correlation functions in the background field formalism, now computed for a macroscopic field $A_\text{bg}[A]$. Establishing the relation $A_\text{bg}[A]$ in practice would be useful for a comparison of methods.

We finally note that gauge invariance and universality strongly restrict the form of gauge-invariant effective actions such as $\bar\Gamma[A]$ or $\bar\Gamma_\text{bg}[A]$. This suggests that $\bar\Gamma[A]$ and $\bar\Gamma_\text{bg}[A]$ may actually be identical up to a non-linear field redefinition, and perhaps up to the value of the gauge coupling at a given momentum or the associated confinement scale.

\section{Gauge-invariant flow equation}
\label{sec:gauge-invariant flow}

The construction of the gauge-invariant effective average action $\Gamma_k[A]$ proceeds by introducing an infrared cutoff function $R_k$ which suppresses the contributions of fluctuations for which $\mathcal{D} \lesssim k^2$ for a suitable generalization $\mathcal{D}$ of the covariant Laplacian. For $k \to 0$ the effective average action $\Gamma_k[A]$ becomes the gauge-invariant quantum effective action $\bar\Gamma[A]$ discussed in the previous section. The dependence of $\Gamma_{k}$ on $k$ obeys a functional flow equation. If $\Gamma_{k}[A]$ is gauge invariant for all $k$, the flow equation has to be gauge invariant. We aim for a closed form for which the flow generator can be computed from $\Gamma_{k}[A]$. The gauge invariant flow equation \eqref{eqn:I1} has been discussed in detail in ref.~\cite{CWGIF} and we only sketch here briefly some of the relevant ingredients. We rather focus on the explicit equations for pure Yang-Mills theories and on the practical use for the computation of the running gauge coupling and the gluon propagator.

While the non-local projectors are important for the general formulation of the gauge invariant flow equation, they do not appear any longer in the explicit form of the flow equation for the configurations considered here. The role of the projectors is to divide the contributions to the flow into two subsectors. In each subsector the flow equations take rather familiar forms, without the appearance of explicit projectors. 

For $k \neq 0$ two modifications are needed in order to obtain a simple closed gauge-invariant flow equation \cite{CWGIF}. First, the macroscopic gauge field $A$ no longer equals the expectation value $\langle A^\prime\rangle$ even for an optimal physical gauge fixing. While $\langle c^\prime\rangle = 0$ remains preserved for $\alpha \to 0$, one now has a $k$-dependent non-zero value $b_\mu = \langle b_\mu^\prime\rangle = \langle A_\mu^\prime\rangle - \hat{A}_\mu$, which is, in principle, computable for any given source $J^\mu$, cf. eq.~\eqref{eqn:G4}. For $k \to 0$ one recovers $\langle b_\mu^\prime\rangle$ as determined by eq.~\eqref{eqn:114D}. For an optimal physical gauge fixing $\langle b^\prime\rangle$ vanishes for $k \to 0$.

Inversely, for any given $b_\mu$ one can determine the associated source $J^\mu$. We may choose for any given $k$ a value $b(k)$ such that the correlation function for $b^\prime$ equals the physical propagator, $\tilde{G} = G$. This modifies the relation \eqref{eqn:96A}, and therefore the implicit definition \eqref{eqn:346A} of the effective action $\tilde\Gamma$. For practical computations the explicit relation between $\langle b_\mu^\prime\rangle$ and the source is not needed. Second, the effective average action $\Gamma_k[A]$ involves a $k$-dependent correction term $C_k[A]$. For $k \to 0$ this equals $C[A]$ as discussed in the previous section. Both $\langle b_\mu^\prime\rangle$ and $C_k$ are determined by the requirement that the flow of $\Gamma_k[A]$ remains simple, computable in terms of $\Gamma_k[A]$ and its functional derivatives.

For comparison we discuss in app.~\ref{app:gauge-invariance flow} a closed flow equation for the case where the source is fixed by eq.~\eqref{eqn:96A} and no correction $C_k$ is added. This again results in a closed gauge-invariant flow equation for a gauge-invariant effective action $\Gamma_k^\prime[A]$. As compared to eq.~\eqref{eqn:I1} it contains correction terms since the identification \eqref{eqn:CF6} of the propagator $G$ and the correlation function $\tilde{G}$ is modified. While the correction terms are, in principle, computable from $\bar\Gamma[A]$ and its functional derivatives, it seems advantageous to employ the freedom in the relation between $b$ and the sources, as well as for the precise definition of $\Gamma_k$, in order to implement the simple form \eqref{eqn:I1} of the flow equation. In this section we perform simple computations of running gauge couplings and propagators for the flow equation \eqref{eqn:I1}. This illustrates how the gauge-invariant flow equation \eqref{eqn:I1} can be used in practice.

\subsection{Flow in pure gauge theories}

As a demonstration of the use of the gauge-invariant flow equation we consider pure non-abelian gauge theories with the simple truncation \eqref{eqn:YX10} for the gauge-invariant effective action. The only flowing parameter is the $k$-dependent gauge coupling $g(k)$. In order to compute the $\beta$-function for $g$ we evaluate the flow equation for configurations obeying
\begin{equation}\label{eqn:F1a}
	F^{\mu\nu}{_{;\nu}}
	= 0.
\end{equation}
In this case we can employ the identities \eqref{eqn:295B}, \eqref{eqn:295C} in order to establish that the projected inverse propagator $\bar\Gamma^{(2)}_P$ equals $\bar{\Gamma}^{(2)}$ in eq.~\eqref{eqn:A21-2}.

Since $\bar{\Gamma}^{(2)}_P$ is a matrix in the space of adjoint indices $z$ we represent the gauge fields here as matrices in the adjoint representation,
\begin{alignedeqn}\label{eqn:F2a}
	(A_\mu)_{yz}
	&= A^w_\mu(T_w)_{yz},
	(F_{\mu\nu})_{yz}
	= F^w_{\mu\nu}(T_w)_{yz},\\
	(T_w)_{yz}
	&= -if_{wyz},
	\qquad
	[T_w,T_y]
	= i \tensor{f}{_w_y^z} T_z.
\end{alignedeqn}
Many formulae of the previous discussion with matrices in the fundamental representation generalize to the adjoint representation. The trace $\Tr$ has to be replaced by a normalized trace $\tilde\Tr$, such that $\tilde\Tr = \Tr$ in the fundamental representation, and $\tilde\Tr = (1/2 N) \, \Tr$ in the adjoint representation of $SU(N)$.

The action of covariant derivatives on the vector field $B^z$ in the adjoint representation, as given by eq.~\eqref{eqn:26A}, \eqref{eqn:23AA}, involves a matrix multiplication according to
\begin{equation}\label{eqn:F3a}
	D_\mu
	= \partial_\mu - i A_\mu.
\end{equation}
We will in the following use this representation of $D_\mu$. If indices are not indicated explicitly, matrices in this section have adjoint indices $z$, $y$. The (projected) second functional derivative of the effective action \eqref{eqn:YX10} (for Minkowski signature) reads
\begin{alignedeqn}\label{eqn:F4a}
	\tensor{(\bar{\Gamma}^{(2)}_P)}{_\mu^\nu}
	&= \frac{i}{g^2} \tensor{\mathcal{D}}{_\mu^\nu},\\
	\tensor{\mathcal{D}}{_\mu^\nu}
	&= \{-D^2\delta^\nu_\mu + D_\mu D^\nu + 2i \tensor{F}{_\mu^\nu}\}.
\end{alignedeqn}

The effective average action or flowing action obtains by adding to the classical action an infrared cutoff piece $\Delta_{k}S$ which suppresses the small (covariant) momentum fluctuations. For the infrared cutoff for the physical fluctuations we choose
\begin{equation}\label{eqn:F5a}
	\Delta_k S
	= \frac{i}{2 g^2} \int_x (A_z^{\prime\mu} - \hat{A}_z^\mu) \, r_{\mu y}^{z \nu}(\mathcal{D}) \, (A_\nu^{\prime y} - \hat{A}_\nu^y),
\end{equation}
with $\tensor{r}{_\mu^\nu}(\mathcal{D})$ a matrix valued function of the operator $\mathcal{D}$. The choice \eqref{eqn:F5a} results in an IR regulated inverse propagator
\begin{equation}\label{eqn:F6a}
	\tilde\Gamma^{(2)}_k
	= \frac{i}{g^2}\bigl(\mathcal{D} + r_k(\mathcal{D})\bigr)
	= \frac{i}{g^2}
	\mathcal{P}_k(\mathcal{D}),
\end{equation}
where we note that the gauge-fixing part of the IR cutoff is not yet included in this definition. (The use of the tilde for $\tilde{\Gamma}_{k}^{(2)}$ follows historical conventions and should not be confounded with the effective action $\tilde\Gamma$ in presence of the gauge-fixing term.) We will require
\begin{equation}\label{eqn:F7a}
	\lim_{x\to 0} r_k(x)
	= k^2,
	\qquad
	\lim_{x\to \infty} r_k(x)
	= 0,
\end{equation}
such that $r_k$ induces an infrared cutoff for eigenvalues of $\mathcal{D}$ smaller than $k^2$, and is ineffective for eigenvalues of $\mathcal{D}$ larger than $k^2$.

For a discussion of the resulting gauge invariant flow equation we refer to \cite{CWGIF}. The contribution from the physical fluctuations reads
\begin{align}\label{eq:AS:145AA} 
k\partial_{k}\Gamma_{k}=\pi_{k}+\ldots\, ,\\
\pi_{k}=\dfrac{1}{2}\tr\left \{k\partial_{k}\tilde{\Gamma}_{k}^{(2)}(P\tilde{\Gamma}_{k}^{(2)}P)^{-1}\right \}\, ,
\end{align}
where the $k$-derivative on the r.h.s. only acts on $r_{k}/g^{2}$. For our purposes eq.~\eqref{eqn:F6a} can be taken as the definition of $\tilde\Gamma^{(2)}_k$, e.g.
\begin{equation}
	\tilde\Gamma^{(2)}_k
	= \bar\Gamma^{(2)} + \frac{i}{g^2} \, r_k(\mathcal{D}).
\end{equation}
In eq.~\eqref{eqn:F6a} $\tilde \Gamma^{(2)}_k$ stands for $\tensor{(\tilde\Gamma^{(2)}_k)}{_\mu^\nu}$, e.g. we have lowered one of the indices. As a result we have
\begin{equation}\label{eqn:F8a}
	\tilde\Gamma^{(2)}_{k,P}
	= P \tilde\Gamma^{(2)}_k P
	= \tilde\Gamma^{(2)}_k,
\end{equation}
where we use the property
\begin{equation}\label{eqn:F9a}
	P \mathcal{D}
	= \mathcal{D} P
	= \mathcal{D},
\end{equation}
that holds for $\tensor{F}{_\mu_\nu_;^\nu} = 0$.

We have to solve the inversion problem
\begin{equation}\label{eqn:F10a}
	\tilde\Gamma^{(2)}_k \tilde{G}_k
	= P,
\end{equation}
for $\tensor{(\tilde{G}_k)}{_\mu^\nu} = (G_P)_{\mu\rho} \, \eta^{\rho\nu}$. For this purpose we employ the decomposition \cite{RW}
\begin{equation}\label{eqn:F11a}
	\mathcal{D}
	= \bar{\mathcal{D}} - \mathcal{D}_L,
\end{equation}
with
\begin{alignedeqn}\label{eqn:F12a}
	\tensor{(\mathcal{D}_L)}{_\mu^\nu}
	&= -D_\mu D^\nu,\\
	\tensor{\bar{\mathcal{D}}}{_\mu^\nu}
	&= -D^2\delta^\nu_\mu + 2i \tensor{F}{_\mu^\nu}.
\end{alignedeqn}
The operator $\bar{\mathcal{D}}$ is invertible on the full function space. The operators $\bar{\mathcal{D}}$ and $\mathcal{D}_L$ commute
\begin{equation}\label{eqn:F13a}
	\bar{\mathcal{D}} \mathcal{D}_L
	= \mathcal{D}_L \bar{\mathcal{D}}
	= \mathcal{D}^2_L,
\end{equation}
and one has
\begin{equation}\label{eqn:F14a}
	\mathcal{D}
	= P \bar{\mathcal{D}}
	= \bar{\mathcal{D}} P,
	\qquad
	\mathcal{D} \mathcal{D}_L
	= \mathcal{D}_L\mathcal{D}
	= 0.
\end{equation}

For any function $f(x)$ that admits a Taylor expansion one finds
\begin{equation}\label{eqn:F15a}
	f(\mathcal{D})
	= f(\bar{\mathcal{D}}) - f(\mathcal{D}_L)
	= P f(\bar{\mathcal{D}}).
\end{equation}
The solution of
\begin{equation}\label{eqn:F16a}
	f(\mathcal{D}) G
	= P
\end{equation}
reads therefore
\begin{equation}\label{eqn:F17a}
	G
	= f^{-1}(\bar{\mathcal{D}}) P
	= f^{-1}(\mathcal{D}).
\end{equation}
This yields
\begin{equation}\label{eqn:F18a}
	\tilde{G}_k
	= -i g^2 \mathcal{P}^{-1}_k(\mathcal{D}).
\end{equation}
In short, the operator $\bar{\mathcal{D}}$ is block diagonal, with $\mathcal{D}$ and $\mathcal{D}_L$ the submatrices in the respective projected spaces. After inversion of the inverse propagator in the appropriate projected subspace the projectors are no longer present in the flow equation.

For the contribution from the physical fluctuations we obtain
\begin{alignedeqn}\label{eqn:F20a}
	\pi_k
	&= \frac{1}{2}\tr\bigl\{\bigl(k\partial_k r_k(\mathcal{D}) - \eta_Fr_k(\mathcal{D})\bigr)\mathcal{P}^{-1}_k(\mathcal{D})\bigr\}\\
	&= \frac{1}{2}\tr\mathcal{H}(\mathcal{D}),
\end{alignedeqn}
with
\begin{equation}\label{eqn:F21a}
	\eta_F
	= \frac{\partial\ln g^2}{\partial\ln k}.
\end{equation}
The function $\mathcal{H}(\mathcal{D})$ reads
\begin{equation}\label{eqn:400A}
	\mathcal{H}(\mathcal{D})
	= (k\partial_k - \eta_F)r_k(\mathcal{D})\bigl(\mathcal{D} + r_k(\mathcal{D})\bigr)^{-1}.
\end{equation}
We finally employ
\begin{equation}\label{eqn:F22a}
	\tr f(\mathcal{D}_L)
	= \tr f(\mathcal{D}_S),
	\qquad
	\mathcal{D}_S
	= -D^2,
\end{equation}
and arrive with eq.~\eqref{eqn:F15a} at
\begin{equation}\label{eqn:F23a}
	\pi_k
	= \frac{1}{2} \tr\mathcal{H}(\bar{\mathcal{D}}) - \frac{1}{2}\tr\mathcal{H}(\mathcal{D}_S).
\end{equation}

The flow equation \eqref{eqn:I1} contains a contribution from physical fluctuations $\pi_k$, corresponding to the first term in eq.~\eqref{eqn:I1}, from gauge fluctuations $\delta_k$, and from the regularization of the Faddeev-Popov determinant $\epsilon_k$,
\begin{equation}\label{eqn:F19a}
	k \partial_k \Gamma_k
	= \zeta_k
	= \pi_k + \delta_k - \epsilon_k.
\end{equation}
For our regularization one has $\epsilon_k = -2 \delta_k$, in accordance with eq.~\eqref{eqn:I1}. Indeed, the regulator term \eqref{eqn:F5a} provides for an infrared cutoff for the transversal fluctuations, but not yet for the gauge fluctuations and the Faddeev-Popov determinant. This is easily seen by computing $\Gamma_k$ for very large $k$. This object is needed as an initial value for the flow and should be sufficiently simple. If the longitudinal fluctuations are not regulated the Faddeev-Popov determinant $M$ and the unregulated gauge fixing term would induce highly complicated non-local terms in $\Gamma_k$. We have therefore to extend the regularization and add the ``measure terms'' $\delta_k$ and $\epsilon_k$ in the flow equation \eqref{eqn:F19a}.

For the gauge fluctuations we introduce an additional cutoff term
\begin{equation}\label{eqn:F24a}
	\Delta_k S_c
	= \frac{1}{2 \alpha} \int_x c_z^{\prime\mu} \, r_k(\mathcal{D}_L)_{\mu y}^{z \nu} \, c_\nu^{^\prime y}.
\end{equation}
Correspondingly, we subtract for the effective action
\begin{equation}\label{eqn:404A}
	\Delta_k\Gamma_\text{gf}
	= \frac{1}{2 \alpha} \int_x \hat{c}_z^\mu \, r_k(\mathcal{D}_L)_{\mu y}^{z \nu} \, \hat{c}_\nu^y.
\end{equation}
Combining eq.~\eqref{eqn:347B} with the second functional derivative of eq.~\eqref{eqn:404A} at $\hat{c} = 0$ one finds
\begin{equation}\label{eqn:F25a}
	\tilde\Gamma^{(2)}_{k,\text{gf}}
	= \frac{1}{\alpha}\bigl((\mathcal{D}_L + r_k(\mathcal{D}_L)\bigr).
\end{equation}
Insertion into the general exact flow equation \cite{CWFE} yields the contribution from the gauge fluctuations
\begin{equation}\label{eqn:405A}
	\delta_k
	= \frac{1}{2}\tr\tilde{\mathcal{H}}(\mathcal{D}_L)
	= \frac{1}{2}\tr\tilde{\mathcal{H}}(\mathcal{D}_S),
\end{equation}
with
\begin{equation}\label{eqn:405B}
	\tilde{\mathcal{H}}(\mathcal{D}_L)
	= k\partial_k r_k(\mathcal{D}_L)\bigl(\mathcal{D}_L + r_k(\mathcal{D}_L)\bigr)^{-1}.
\end{equation}
As compared to $\mathcal{H}(\mathcal{D})$ in eq.~\eqref{eqn:400A} the contribution $\sim \eta_F$ is absent. For $\eta_F = 0$ the measure contribution $\delta_k$ cancels the second term in $\pi_k$ in eq.~\eqref{eqn:F23a}.

The Faddeev-Popov determinant takes the form
\begin{equation}\label{eqn:449A}
	M
	= \Det\bigl[\mathcal{D}_S + iD^\mu(A_\mu^\prime - \hat{A}_\mu)\bigr].
\end{equation}
Without an additional regularization this would produce for $k\rightarrow\infty$ a complicated term in the effective action. We want to introduce a regulator that guarantees simplicity of $\Gamma_{k\rightarrow\infty},$ while it becomes absent for $k\rightarrow 0$. We therefore insert into the functional integral a regulator factor \cite{RW}
\begin{equation}\label{eqn:405C}
	E_k
	= \frac{\Det\bigl(\mathcal{D}_S + r_k(\mathcal{D}_S)\bigr)}{\Det\bigl(\mathcal{D}_S\bigr)}.
\end{equation}
It becomes unity for $k = 0$ and therefore ineffective. In the presence of the regulator we replace $M$ by
\begin{equation}\label{eqn:451}
	E_k M
	= \Det\bigl(\mathcal{D}_S + r_k(\mathcal{D}_S)\bigr){\det}\bigl(1 + i\mathcal{D}_S^{-1}D^\mu(A_\mu^\prime - \hat{A}_\mu)\bigr)
\end{equation}

In the limit $k\to \infty$ the regulator function $r_k\approx k^2$ dominates such that $E_kM$ becomes an irrelevant field independent constant,
\begin{alignedeqn}\label{eqn:405E}
	\lim_{k \to \infty} E_k M
	&= \Det(k^2)\Det\left(1 + i\mathcal{D}^{-1}_S D^\mu(A_\mu^\prime - \hat{A}_\mu \right)\\
	&\approx \Det(k^2).
\end{alignedeqn}
Indeed, for $k\rightarrow\infty$ the functional integral for $A^\prime$ contains now diverging quadratic terms for all fluctuations $A^\prime - \hat{A}$. The saddle point approximation becomes exact and we can replace in eq.~\eqref{eqn:451} $A^\prime\to \hat{A}$. Generalizing for finite but very large $k$ we conclude that
$\Gamma_k$ becomes indeed simple. If we express $E_kM$ in terms of ghost fields the term $\sim r_k$ regulates the ghost propagator.

The $k$-dependence of $E_k$ arises only through $r_k$ and we infer the measure contribution from the regularization of the Faddeev-Popov determinant
\begin{alignedeqn}\label{eqn:452A}
	\epsilon_k
	&= k\partial_k \ln E_k\\
	&= k\partial_k \tr\ln \bigl(\mathcal{D}_S + r_k(\mathcal{D}_S)\bigr)\\
	&= \tr\bigl\{k\partial_k r_k(\mathcal{D}_S)(\mathcal{D}_S + r_k(\mathcal{D}_S))^{-1}\bigr\}\\
	&= \tr\tilde{\mathcal{H}}(\mathcal{D}_S)
	= 2 \delta_k.
\end{alignedeqn}
We therefore end with
\begin{equation}\label{eqn:452B}
	k \partial_k \Gamma
	= \pi_k - \delta_k.
\end{equation}
At this stage $\delta_{k}$ is a functional of $\hat{A}(A)$. The extension to a gauge invariant functional of $A$ is straightforward. Since $\mathcal{D}_{S}$ in eq.~\eqref{eqn:405A} involves only covariant derivatives we can simply evaluate eq.~\eqref{eqn:405A} for arbitrary $A$. This yields already a gauge invariant expression.

The regularization \eqref{eqn:405C} has the advantage that the combined ``measure term''-$\delta_k$ is a fixed functional of $A_\mu$, not involving the form of the effective average action $\Gamma_k[A]$. For a given choice of $r_k$ the trace \eqref{eqn:405A} can be computed independent of the truncation for $\Gamma_k$. An alternative regularization could employ a formulation with ghosts and introduce an explicit IR regularization for the ghost propagator. In this case one would have to follow the flow of the combined effective action for gauge fields and ghosts. For the approximations employed in the present paper the two alternatives give identical results. It remains to be seen in practice which choice of regularization of the Faddeev-Popov determinant is best for precision computations.

We observe that the relation $\delta_k - \epsilon_k = -\delta_k = -\epsilon_k/2$ is no accident. For $\alpha \to 0$ the longitudinal sector decouples from the physical sector. In the absence of an IR cutoff in the longitudinal sector the integration over the longitudinal sector would produce a factor $\bigl(\det[\mathcal{D}_S(A)]\bigr)^{-1/2}$. Instead of the IR regularization \eqref{eqn:F24a} we could insert an overall regularization factor $E_k^{-1/2}$, cf. eq.~\eqref{eqn:405C}, similar to the regularization of the Faddeev-Popov determinant. This replaces $\det(\mathcal{D}_S)^{-1/2}$ by the regularized expression $\det[\mathcal{D}_S + r_k(\mathcal{D}_S)]^{-1/2}$. The total measure factor for longitudinal fluctuations and Faddeev-Popov determinant amounts then to $E_k^{1/2}$, resulting in a total measure contribution to the flow equation of $-\epsilon_k/2 = -\delta_k$. The same total measure factor $E_k^{-1/2}$ arises if we adopt the physical gauge fixing \eqref{eqn:76A}. In this case the Faddeev-Popov determinant needs no regularization. On the other hand, the unregularized integral over the longitudinal fluctuations would now produce a factor $\det(\mathcal{D}_S(A))^{-1/2}$. This can be regularized by a factor $E_k^{-1/2}$.

\subsection{Running coupling in \texorpdfstring{$SU(N)$}{SU(N)}-gauge theories}

We specialize to the non-abelian gauge group $SU(N)$ where
\begin{equation}\label{eqn:B1}
	\Tr(T_y T_z)
	= N \delta_{yz},
\end{equation}
such that
\begin{equation}\label{eqn:B2}
	F^z_{\mu\nu} F^{\mu\nu}_z
	= \frac{1}{N} \Tr F_{\mu\nu} F^{\mu\nu}
	= 2 \tilde\Tr F_{\mu\nu} F^{\mu\nu}.
\end{equation}
For the computation of the flow equation for the gauge coupling $g$ we evaluate eq.~\eqref{eqn:452B} for a configuration that corresponds to a constant color-magnetic field
\begin{equation}\label{eqn:B3}
	A^z_\mu(x)
	= n^z A^{(B)}_\mu(x),
	\qquad
	A_\mu(x)
	= A^{(B)}_\mu(x) n_z T^z,
\end{equation}
with $A^{(B)}_\mu(x)$ an abelian gauge field corresponding to a constant magnetic field $B$, e.g. $\partial_1 A^{(B)}_2 - \partial_2 A^{(B)}_1 = B$, and
\begin{equation}\label{eqn:B4}
	F^{\mu\nu}_z F^z_{\mu\nu}
	= 2 B^2.
\end{equation}
Choosing a transversal gauge field, $\partial^{\mu}A_{\mu}^{(B)}=0$, the configuration \eqref{eqn:B3} is a physical gauge field of the type \eqref{eq:AS:60A}.

We perform the computation in Euclidean space and analytically continue later to Minkowski space. For Euclidean signature the term in the effective action quadratic in $B$ is identified with eq.~\eqref{eqn:YX10}
\begin{equation}\label{eqn:B5}
	\bar\Gamma(B)
	= \int_x \frac{B^2}{2 g^2} + \dots
\end{equation}
The flow of the gauge coupling can therefore be extracted as
\begin{equation}\label{eqn:B6}
	k \partial_k \left(\frac{1}{g^2}\right)
	= \Omega^{-1}
	\frac{\partial^2}{\partial B^2} k \partial_k
	\bar\Gamma(B)|_{B=0},
\end{equation}
with total volume $\Omega = \int_x$.

For the evaluation of the traces \eqref{eqn:F23a}, \eqref{eqn:405A} we can closely follow ref.~\cite{RW}. One has in quadratic order in $B$
\begin{equation}\label{eqn:B7}
	\Omega^{-1}\tr\mathcal{H}(\bar{\mathcal{D}})\Rightarrow
	\frac{5N}{24\pi^2}\mathcal{H}(\mathcal{D}=0) B^2
	= \frac{5(2 - \eta_F)NB^2}{24\pi^2},
\end{equation}
where we employ
\begin{equation}\label{eqn:B8}
	\mathcal{H}(\mathcal{D}=0)
	= 2 - \eta_F.
\end{equation}
Similarly, one finds for the terms $\sim B^2$
\begin{alignedeqn}\label{eqn:B9}
	&\Omega^{-1}\tr\mathcal{H}(\mathcal{D}_S)
	\Rightarrow -\frac{(2 - \eta_F)NB^2}{96\pi^2},\\
	&\Omega^{-1}\tr\tilde{\mathcal{H}}(\mathcal{D}_S)
	\Rightarrow -\frac{NB^2}{48\pi^2},
\end{alignedeqn}
such that
\begin{alignedeqn}\label{eqn:B10}
	&\pi_k
	= \frac{21(2 - \eta_F)N\Omega B^2}{192\pi^2},
	\quad
	\delta_k
	= -\frac{N\Omega B^2}{96\pi^2},\\
	&\pi_k - \delta_k
	= \frac{N\Omega B^2}{48\pi^2}
	\left(11 - \frac{21}{4}\eta_F\right).
\end{alignedeqn}

One arrives at a non-linear equation for $\eta_F$
\begin{equation}\label{eqn:B10a}
	k\frac{\partial}{\partial k}
	\left(\frac{1}{g^2}\right)
	= \frac{N}{24\pi^2}
	\left(11 - \frac{21}{4}\eta_F\right)
	= -\frac{\eta_F}{g^2},
\end{equation}
that is solved by
\begin{equation}\label{eqn:B11}
	\eta_F
	= -\frac{11 Ng^2}{24 \pi^2}
	\left(1 - \frac{7 Ng^2}{32 \pi^2}\right)^{-1}.
\end{equation}
Our truncation therefore yields for the flow of the gauge coupling
\begin{alignedeqn}\label{eqn:B12}
	\partial_tg^2
	&= \eta_Fg^2
	= -\frac{11 Ng^4}{24 \pi^2}
	\left(1 - \frac{7 Ng^2}{32 \pi^2}\right)^{-1}\\
	&= -\frac{11 Ng^4}{24 \pi^2} - \frac{77 N^2 g^6}{768 \pi^4} - \dots
\end{alignedeqn}
The first term is the usual one-loop beta-function, while the second almost reproduces the exact two-loop contribution, for which the factor $77$ has to be replaced by $68$.

\subsection{Flow of the gauge field propagator}

The propagator $G_k$ for the physical fluctuations of gauge fields is the inverse of the second functional derivative of $\Gamma_k$ on the projected subspace of physical fluctuations,
\begin{equation}
	\Gamma_k^{(2)} \, G_k
	= P.
\end{equation}
The inverse propagator $\Gamma_k^{(2)}$ is related to $\tilde\Gamma_k^{(2)}$ in eq.~\eqref{eqn:F6a} by subtraction of the IR cutoff piece, and reads in our truncation
\begin{equation}\label{eqn:F1}
	\Gamma_k^{(2)}
	= \frac{i \, \mathcal{D}}{g^2}.
\end{equation}
Its flow is given by the $k$-dependence of $g$, e.g.
\begin{equation}\label{eqn:F2}
	\partial_t \Gamma_k^{(2)}
	= -\eta_F \, \Gamma_k^{(2)}.
\end{equation}
Since the projector $P$ does not depend on $k$, this transfers directly to
\begin{equation}\label{eqn:F3}
	\partial_t G_k
	= \eta_F \, G_k.
\end{equation}
We observe that for the gauge-invariant formulation of the flow equation the anomalous dimension $\eta_F$ is the same object that appears in the $\beta$-function \eqref{eqn:F21a} for the running gauge coupling. This contrasts with formulations with a fixed background field where $\eta_F$ is replaced to lowest order by $\tilde\eta_F = (13/22) \eta_F$. We will discuss the origin of this difference below.

An improved truncation for the (Euclidean) effective average action is given by
\begin{equation}\label{eqn:F4}
	\Gamma_k
	= \frac{1}{4} \int_x \tensor{F}{_z_\mu^\nu} \, Z_{\nu y}^{z \rho}(k^2 + \mathcal{D}) \, F_\rho^{y \mu},
\end{equation}
with $\tensor{Z}{_\nu^\rho}(x)$ given for $x = k^2$ by a solution of the flow equation for $g^{-2}$,
\begin{equation}\label{eqn:F5}
	Z(x = k^2)
	= g^{-2}(k^2).
\end{equation}
Since $\mathcal{D}$ is matrix valued, also $Z$ is matrix valued. In perturbation theory for small $g^2$ the improvement is a higher-order effect. To lowest order one has
\begin{equation}\label{eqn:F6}
	Z
	= \frac{1}{g^2(k_0)} + \frac{11 \, N}{48 \pi^2} \, \ln\biggl(\frac{k^2 + \mathcal{D}}{k_0^2}\biggr).
\end{equation}
For $\mathcal{D} \ll k^2$ this reduces to the truncation \eqref{eqn:YX10}, while for $\mathcal{D} \gg k^2$ we take into account that external momenta or large fields act as physical infrared cutoffs that effectively stop the flow. At quadratic order in $A_\mu$ one can replace
\begin{equation}
	\tensor{\mathcal{D}}{_\mu^\nu}
	\to -\partial^2 \, \delta_\mu^\nu + \partial_\mu \partial^\nu.
\end{equation}

We can repeat the computation \eqref{eqn:F20a} of $\pi_k$, with $\mathcal{H}(\mathcal{D})$ replaced by
\begin{equation}\label{eqn:F7}
	\mathcal{H}^\prime(\mathcal{D})
	= (\partial_t - \eta_F) \, r_k \, \bigl[g^2 \, Z(\mathcal{D}) \, \mathcal{D} + r_k(\mathcal{D})\bigr]^{-1}.
\end{equation}
If we define the running gauge coupling as before from the zero momentum limit of the effective action, the result for the $\beta$-function remains the same since $\mathcal{H}^\prime(0) = \mathcal{H}(0)$. For an alternative definition at non-zero momentum, $p^2 \neq 0$, the running is effectively stopped for $k^2 < p^2$. Taking the limit $k \to 0$ the gauge coupling becomes a function of $p^2$, with dependence on $p^2$ governed by the same $\beta$-functions as for the dependence on $k^2$ for $p^2 \ll k^2$. For the Euclidean inverse gluon propagator the improved truncation yields a non-trivial momentum dependence according to
\begin{equation}\label{eqn:F8}
	\Gamma^{(2)}
	= Z(k^2 + \mathcal{D}) \, \mathcal{D}.
\end{equation}

Quite generally, the gauge field propagator (in the ``vacuum state'' $A_\mu = 0$) is expected to be a function of $\mathcal{D}$, and we may write
\begin{alignedeqn}\label{eqn:F9}
	\Gamma_{k,zy}^{\mu\nu}(x,y)
	&= \frac{\partial^2 \Gamma_k}{\partial A_\mu^z(x) \, \partial A_\nu^y(y)}\\
	&= \delta(x - y) \, \eta^{\mu\rho} \, \tensor{\bigl[Z(k^2,\mathcal{D}) \, \mathcal{D}\bigr]}{_z_y_\rho^\nu},
\end{alignedeqn}
extending eq.~\eqref{eqn:F8} to a more general form of $Z$. In eq.~\eqref{eqn:F9} all quantities are evaluated for $A_\mu = 0$. The ansatz \eqref{eqn:F8} is compatible with the gauge invariance of $\Gamma_k$ and the projection onto physical fluctuations. The factor $\delta(x - y)$ reflects translation symmetry, with $\mathcal{D}$ acting on $y$.

The flow equation for $\Gamma_k^{(2)}$ is obtained from eq.~\eqref{eqn:I1} or \eqref{eqn:F19a} by taking two derivatives with respect to the gauge fields
\begin{equation}\label{eqn:F10}
	\partial_t {\Gamma_k^{(2)}}_{zy}^{\mu\nu}(x,y)
	= \frac{\partial^2}{\partial A_\mu^z(x) \, \partial A_\nu^y(y)} \, \bigl(\pi_k - \delta_k\bigr).
\end{equation}
In contrast to formulations with a fixed background field also $r_k$ in eq.~\eqref{eqn:F20a}, \eqref{eqn:400A} depends on the macroscopic gauge field, such that eq.~\eqref{eqn:F10} receives contributions from $\partial r_k/\partial A_\mu^z$, etc. (This holds analogously for the measure contribution $\delta_k$.) One can cast eq.~\eqref{eqn:F10} into a sum of one-loop diagrams with two external legs and one insertion of $\partial_t R_k$. For a fixed background field this involves the usual three- and four-point vertices, obtained from third and fourth derivatives of $\Gamma_k$. For our gauge-invariant formulation one has additional diagrams involving derivatives of $r_k$ with respect to the gauge field.

In the truncation \eqref{eqn:YX10} the flow of the inverse propagator at zero momentum is given by eq.~\eqref{eqn:F2}. The difference to the result for a fixed background field arises precisely from the terms $\sim \partial r_k/\partial A_\mu$. For non-zero momentum squared $p^2 \neq 0$ one expects that the external momentum stops the flow for $k^2 < p^2$. This results for $Z$ in the qualitative behavior \eqref{eqn:F5}. A more precise estimate of $Z$ needs an explicit computation.

Extrapolating our estimate for $g(k)$ in the truncation \eqref{eqn:F4}, \eqref{eqn:F5} to small $k$ would lead to vanishing of $Z$ for small momenta. This clearly indicates the insufficiency of this truncation for the infrared behavior of Yang-Mills theories. A self-consistent flow for the truncation
\begin{equation}\label{eqn:191B}
	\Gamma
	= \frac{1}{4} \int_x \tensor{F}{_z_\mu^\nu} \, Z_{\nu y}^{z \rho}(\mathcal{D}) \, F_\rho^{y \mu}
\end{equation}
would be interesting. As long as the ground state corresponds to $A_\mu = 0$ eq.~\eqref{eqn:191B} accounts for the propagator of the gauge fields. In QCD the Fourier transform of $\mathcal{D} Z(\mathcal{D})$ (at $A = 0$) is related to the heavy quark potential. A behavior $Z \sim \mathcal{D}$ for $k \to 0$ corresponds to a linearly rising potential in position space.

\section{Conclusions}
\label{sec:conclusions}

In this paper we propose a gauge-invariant effective action for theories with local gauge symmetry. This effective action depends on only one macroscopic gauge field $A_\mu$, in distinction to the background field formalism. The macroscopic gauge field appears in the functional integral defining the effective action -- this turns the definition formally into a functional differential equation. In practice, our construction proceeds by a particular physical gauge fixing, determined such that the gauge-fixed effective action $\tilde\Gamma[A]$ decays into a gauge-invariant physical part $\bar\Gamma[A]$ and a gauge part $\Gamma_\text{gf}[A]$,
\begin{equation}\label{eqn:ZA}
	\tilde\Gamma[A]
	= \bar\Gamma[A] + \Gamma_\text{gf}[A].
\end{equation}

Arbitrary gauge fields can be split into physical gauge fields $\hat{A}$ and gauge degrees of freedom $\hat{c}$, $A = \hat{A}(A) + \hat{c}(A)$. Physical gauge fields obey differential constraints. They formally depend on the choice of a reference field, which does not matter in practice, however. The gauge-fixing term is quadratic in $\hat{c}$, with coefficient $\alpha^{-1}$ tending to infinity. Solutions of the field equations for arbitrary (finite) sources imply $\hat{c} = 0$. Inserting $\hat{c} = 0$ into eq.~\eqref{eqn:ZA} projects onto $\bar\Gamma$. The gauge-invariant effective action $\bar\Gamma$ depends on $A$ only via the physical gauge fields $\hat{A}$, $\bar\Gamma[A] = \bar\Gamma\bigl[\hat{A}(A)]$, with $\bar\Gamma[\hat{A}] = \tilde\Gamma[A,\hat{c} = 0]$.

We have shown the following properties of the gauge-invariant effective action:
\begin{enumerate}[label=(\roman*)]
	\item The first functional derivative of $\bar\Gamma$ yields the exact quantum field equations for arbitrary conserved sources.
	\item The macroscopic gauge field $A$ equals the expectation value of the microscopic gauge field $\langle A^\prime\rangle$, as computed for a conserved source corresponding to $\partial \bar\Gamma/\partial A$, only for particular choices of macroscopic field and physical gauge fixing.
	\item The inverse propagator for physical fluctuations is defined as the second functional derivative $\bar\Gamma^{(2)}$ of the gauge-invariant effective action. It can be inverted within the projected function space of physical fluctuations. The inverse of $\bar\Gamma^{(2)}$ equals the connected two-point correlation function for the physical fluctuations only for an optimal choice of macroscopic gauge field, effective action and physical gauge fixing.
\end{enumerate}

For an optimal setting the gauge-invariant effective action can be used in many respects in the same way as for theories without local gauge invariance. We have proposed an optimal physical gauge fixing, but not yet explored its use in practice. For other physical gauge fixings the consequences of the difference between the macroscopic field and the expectation value of the microscopic field, as well as between the propagator and the correlation function, need to be explored. We have given an implicit definition of the effective action $\tilde\Gamma$ by a functional differential equation. For small gauge coupling this can be employed for developing perturbation theory by an iterative solution.

We also introduce a gauge-invariant effective average action $\Gamma_k[A]$ which effectively only includes fluctuations with $\mathcal{D} \geq k^2$, where $\mathcal{D}$ is an appropriate covariant Laplacian-type operator. This effective average action equals $\bar\Gamma[A]$ if the infrared cutoff scale $k$ vanishes. The $k$-dependence of $\Gamma_k$ obeys a closed flow equation which takes a one-loop form. It involves the full field-dependent propagator, as given by the inverse of $\bigl(\bar\Gamma_k^{(2)} + R_k\bigr)$, with $R_k$ an IR cutoff term. 

The advantage of the present formulation of $\Gamma_k[A]$ is the closed form of the flow equation for a gauge-invariant object depending on only one macroscopic gauge field. The flow can be computed in terms of $\Gamma_k[A]$ and its derivatives. In contrast, the flow equation in the background field formalism needs information from $\Gamma_{k,\text{bg}}[A,\bar{A}]$ for $A \neq \bar{A}$ \cite{RW}. As compared to the background field formalism with physical gauge fixing we can view the present formulation as following a different trajectory in the space of actions. For $k = 0$ the result may differ from the gauge-invariant effective action in the background formalism $\Gamma_\text{bg}[A,\bar{A} = A]$ by a non-linear field redefinition of the macroscopic gauge field. The relation between the macroscopic fields in the two formulations may be non-local.

For pure Yang-Mills theories we have computed the running of the gauge coupling by use of the gauge-invariant flow equation in a simple truncation. It parallels the computation in the background field formalism of ref.~\cite{RW} up to small modifications. The simple truncation with $\Gamma_k$ given by eq.~\eqref{eqn:YX10} yields the correct one-loop expression for the $\beta$-function, as well as $5/6$ of the two-loop term and higher corrections. We also have computed the flow of the propagator for physical fluctuations. For $k \neq 0$ it differs from the propagator in the background field formalism by a different wave function renormalization. Based on our result we also propose a simple improvement of the effective action beyond the form eq.~\eqref{eqn:YX10}. It will be interesting to see if the gauge-invariant flow equation beyond its simplest truncation can describe successfully the infrared behavior of Yang-Mills theories.

\paragraph{Acknowledgements} The author would like to thank \href{mailto:j.pawlowski@thphys.uni-heidelberg.de}{J. Pawlowski} for stimulating discussions. This work is supported by DFG Collaborative Research Centre ``\href{http://www.dfg.de/en/research_funding/programmes/list/projectdetails/index.jsp?id=273811115&sort=var_asc&prg=SFB}{SFB 1225} (ISOQUANT)'' and ERC-advanced grant \href{http://cordis.europa.eu/project/rcn/101262_en.html}{290623}.

\begin{appendices}

\section{Non-linear decomposition of gauge fields}
\label{app:non-linear decomposition}

In this appendix we relate the gauge degrees of freedom $\hat{c}_\mu$ or $\hat{c}_\mu^\prime$ to the gauge orbits in a non-linear decomposition of the gauge fields $A_\mu$ or $A_\mu^\prime$.

\subsection{Gauge orbits}

The split $A_\mu = \hat{A}_\mu + \hat{c}_\mu$ into a gauge-invariant physical gauge field $\hat{A}_\mu$ and a gauge degree of freedom $\hat{c}_\mu$ can be associated with a non-linear decomposition of the gauge field,
\begin{equation}\label{eqn:A1}
	A_\mu
	= w \, \hat{A}_\mu \, w^\dagger - i (\partial_\mu w) \, w^\dagger,
	\qquad
	w^\dagger w
	= 1.
\end{equation}
We focus here on a gauge symmetry $SU(N)$, with $A_\mu$ and $w$ $N \times N$-matrices and $\det w = 1$. Gauge transformations leave $\hat{A}_\mu$ invariant and transform
\begin{equation}\label{eqn:A2}
	w
	\to u \, w,
\end{equation}
such that
\begin{equation}\label{eqn:A3}
	A_\mu
	\to u \, A_\mu \, u^\dagger - i (\partial_\mu u) \, u^\dagger.
\end{equation}
The non-linear fields $w$ parametrize the gauge orbits associated to $\hat{A}$, and gauge transformations simply act as matrix multiplications of the $SU(N)$-matrices $u$ and $w$. The gauge degrees of freedom $\hat{c}_\mu$ are related to $w$ by
\begin{alignedeqn}\label{eqn:A4}
	 \hat{c}_\mu
	 &= A_\mu - \hat{A}_\mu
	 = w \, \hat{A}_\mu \, w^\dagger - \hat{A}_\mu - i (\partial_\mu w) \, w^\dagger\\
	 &= -i \bigl(D_\mu(\hat{A}) \, w\bigr) \, w^\dagger,
\end{alignedeqn}
with
\begin{equation}
	D_\mu w
	= \partial_\mu w - i [\hat{A}_\mu,w].
\end{equation}

For infinitesimal gauge transformations one has
\begin{equation}\label{eqn:A5}
	u
	= 1 + i \varphi,
	\qquad
	\varphi^\dagger
	= \varphi,
	\qquad
	\tr\varphi
	= 0,
\end{equation}
and recovers eq.~\eqref{eqn:G1}. Similarly, we may consider $w$ close to one
\begin{equation}\label{eqn:A6}
	w
	= 1 + i \delta,
	\qquad
	\delta^\dagger
	= \delta,
	\qquad
	\tr\delta
	= 0.
\end{equation}
To linear order in $\delta$ eq.~\eqref{eqn:A4} yields
\begin{equation}\label{eqn:A7}
	\hat{c}_\mu
	= \partial_\mu \delta - i [\hat{A}_\mu,\delta]
	= D_\mu(\hat{A}) \, \delta
	= D_\mu(A) \, \delta,
\end{equation}
where we employ that $A_\mu - \hat{A}_\mu$ is of the order $\delta$. The gauge condition $D^\mu \, \hat{c}_\mu = 0$ translates to $D^2 \, \delta = 0$. We note that we could also formulate the gauge fixing with $D^\mu(A) \, \hat{c}_\mu$ instead of $D^\mu[\hat{A}] \, \hat{c}_\mu$. To linear order in $\hat{c}$ this makes no difference.

With the decomposition \eqref{eqn:A1} one finds for the field strength
\begin{equation}\label{eqn:A8}
	F_{\mu\nu}
	= w \bigl(\partial_\mu \hat{A}_\nu - \partial_\nu \hat{A}_\mu - i [\hat{A}_\mu,\hat{A}_\nu]\bigr) w^\dagger.
\end{equation}
Therefore invariants such as $\Tr F_{\mu\nu} F^{\mu\nu}$ do not involve $w$. They depend only on the physical gauge fields $\hat{A}_\mu$ and not on the gauge degrees of freedom $\hat{c}_\mu$. This completes the discussion in sect.~\ref{sec:physical gauge fields}. To linear order $\hat{c}_\mu$ is longitudinal, cf. eq.~\eqref{eqn:A7}.

If we consider $A_\mu^\prime$ in the vicinity of a given gauge-invariant field $A_\mu = \hat{A}_\mu$,
\begin{equation}\label{eqn:A9}
	A_\mu^\prime
	= \hat{A}_\mu + h_\mu
	= \hat{A}_\mu + \hat{h}_\mu + \hat{c}_\mu
	= \hat{A}_\mu + f_\mu + D_\mu \, \lambda,
\end{equation}
we can identify to linear order $f_\mu = \hat{h}_\mu$, $\hat{c}_\mu = D_\mu \, \lambda$, such that $\hat{h}_\mu$ is transversal. Beyond linear order $\hat{c}_\mu$ needs no longer to be longitudinal, and $\hat{h}_\mu$ is not transversal. Expanding $w = \exp(i \delta)$ to quadratic order in $\delta$,
\begin{equation}\label{eqn:A10}
	w
	= 1 + i \delta - \frac{i}{2} \delta^2,
\end{equation}
yields
\begin{equation}\label{eqn:A11}
	\hat{c}_\mu
	= D_\mu(\hat{A}) \, \delta - \frac{i}{2} [D_\mu(\hat{A}) \, \delta,\delta],
\end{equation}
which is, in general, not longitudinal. We have discussed this issue in sect.~\ref{sec:physical gauge fields}.

The non-linear field parametrization \eqref{eqn:A1} permits a close contact to a similar parametrization in work on spontaneous color symmetry breaking \cite{CWSBC,CWHP}. (The identification is $v = w^\transp$, $V_\mu = -\hat{A}_\mu^\transp$.) It is straightforward to extend this parametrization to quark fields or other matter fields. For a field $\psi$ in the fundamental representation of $SU(N)$, such as quarks in QCD or the Higgs doublet for the $SU(2) \times U(1)$ electroweak gauge theory, one employs
\begin{equation}\label{eqn:A12}
	\psi
	= w \hat\psi,
\end{equation}
with $\hat\psi$ a gauge-invariant field. For fermions the gauge-invariant kinetic term depends only on the gauge-invariant fields $\hat\psi$ and $\hat{A}_\mu$, not on $w$,
\begin{equation}\label{eqn:A13}
	i \bar\psi_i \, \gamma^\mu \, \bigl(D_\mu(A)\bigr)_{ij} \, \psi_j
	= i \hat{\bar{\psi_i}} \, \gamma^\mu \, \bigl(D_\mu(\hat{A})\bigr)_{ij} \, \hat\psi_j.
\end{equation}

For three-flavor QCD a simple effective action, based on the invariants \eqref{eqn:YX10}, \eqref{eqn:A13}, together with a part for scalar bilinears $\sim \bar\psi \, \psi$, gives a rather satisfying description of phenomenology provided chiral symmetry breaking occurs also in the octet sector \cite{CWSBC}. The physical gauge fields $\hat{A}_\mu$ can then be associated with the octet of vector mesons, and the physical fermions $\hat\psi$ with an octet and singlet of nucleons.

\subsection{Physical gauge fixing}
\label{app:physical gauge fixing}

The non-linear representation sheds light on the physical gauge fixing. To linear order one has
\begin{equation}\label{eqn:A14}
	\delta
	= D^{-2}(\hat{A}) \, D^\mu(\hat{A}) \, (A_\mu - \hat{A}_\mu)
	= \delta^z \, t_z.
\end{equation}
Employing the general relation \eqref{eqn:26B}, covariant derivatives act by matrix multiplication on the vector field $A_\mu^y$
\begin{equation}\label{eqn:190A}
	\delta^z
	= \tensor{\bigl(D^{-2}(\hat{A}) \, D^\mu(\hat{A})\bigr)}{^z_y} \, (A_\mu^y - \hat{A}_\mu^y).
\end{equation}
One may define similarly
\begin{equation}\label{eqn:A15}
	\delta^\prime
	= D^{-2}(\hat{A}) \, D^\mu(\hat{A}) \, (A_\mu^\prime - \hat{A}_\mu)
	= \delta^{\prime z} \, t_z,
\end{equation}
with
\begin{equation}\label{eqn:191A}
	\delta^{\prime z}
	= \tensor{\bigl(D^{-2}(\hat{A}) \, D^\mu(\hat{A})\bigr)}{^z_y} \, (A^\prime - \hat{A})_\mu^y.
\end{equation}
We define a non-local physical gauge fixing, $\alpha \to 0$,
\begin{equation}\label{eqn:A16}
	S_\text{gf}
	= \frac{1}{2 \alpha} \int_x \sum_z (\delta^{\prime z})^2,
\end{equation}
resulting in
\begin{equation}\label{eqn:A17}
	\Gamma_\text{gf}
	= \frac{1}{2 \alpha} \int_x \sum_z (\delta^z)^2.
\end{equation}
The solution of the field equations in the longitudinal sector directly imply $\delta = 0$, instead of $D^2 \, \delta = 0$. For Landau gauge we have disregarded the discussion of non-trivial solutions of $D^2 \, \delta = 0$ and employed the solution $D_\mu \delta = \hat{c}_\mu = 0$. Then the gauge fixing \eqref{eqn:A16} and Landau gauge yield the same results.

From a conceptual point of view the physical gauge fixing \eqref{eqn:A16} seems to be attractive. The Faddeev-Popov determinant reads now, with covariant derivatives given by eq.~\eqref{eqn:23AA} and $\hat{A}$, $A^\prime$ in the adjoint representation \eqref{eqn:F2a},
\begin{alignedeqn}\label{eqn:A18}
	\tilde{M}
	&= \Det\bigl[D^{-2}(\hat{A}) \, D^\mu(\hat{A}) \, D_\mu(A^\prime)\bigr]\\
	&= \Det\bigl[1 - i D^{-2}(\hat{A}) \, D^\mu(\hat{A}) \, (b_\mu^\prime + c_\mu^\prime)\bigr],
\end{alignedeqn}
where we employ the identity
\begin{alignedeqn}\label{eqn:214A}
	D^2(\hat{A}) - D^\mu(\hat{A}) \, D_\mu(A^\prime)
	&= i D^\mu(\hat{A}) \, (A_\mu^\prime - \hat{A}_\mu)\\
	&= i D^\mu(\hat{A}) \, (b_\mu^\prime + c_\mu^\prime).
\end{alignedeqn}

At lowest order in a saddle point approximation one has $\delta^\prime = 0$, $c_\mu^\prime = 0$, $\hat{c}_\mu = 0$ and we recall that this approximation becomes exact for $\alpha \to 0$. One ends effectively with
\begin{equation}\label{eqn:214B}
	\tilde{M}
	= \Det\bigl[1 - i D^{-2}(\hat{A}) \, D^\mu(\hat{A}) \, T_z \, b_\mu^{\prime z}\bigr].
\end{equation}
For $\alpha \to 0$ no change occurs if we change in the definition of $\delta^\prime$ the covariant derivative $D^\mu(\hat{A}) \to D^\mu(A)$. The dependence of $\tilde{M}$ on the transverse fluctuations $b^\prime$ remains a complication for many practical computations, for example perturbative expansions.

The gauge fixing \eqref{eqn:A16} is not very convenient for perturbation theory. In the longitudinal sector one has $\Gamma_\text{gf}^{(2)} \sim \frac{1}{\alpha} \, D^\mu \, D^{-4} \, D^\nu$. For finite $\alpha$ this leads to a strong suppression of the IR fluctuations, while the inverse propagator vanishes as covariant momenta tend to infinity. For any finite momenta the divergence of $1/\alpha$ overwhelms this effect, but a careful discussion of limits is needed. For functional flow equations, which are UV finite by construction, this issue is solved automatically. For perturbation theory it seems advantageous to stick to Landau gauge fixing.

The gauge fixing \eqref{eqn:A16} is not an optimal physical gauge fixing in the sense that $A_\mu$ does not equal the expectation value $\langle A_\mu^\prime\rangle$, and the second functional derivative of $\bar\Gamma[A]$ does not equal the correlation function for the physical fluctuations. The simplest way to realize ``optimal gauge fixing'' would be, of course, to keep the gauge fixing independent of the macroscopic field. We will next discuss a non-linear gauge fixing of the type \eqref{eqn:A16}, with a different non-linear definition of $\delta^\prime$.

We can associate to every microscopic gauge field $A_\mu^\prime$ the corresponding physical gauge field $\smash{\hat{A}_\mu^\prime[A_\mu^\prime]}$, and define $\delta^\prime$ by
\begin{alignedeqn}\label{eqn:214C}
	A_\mu^\prime
	&= e^{i \delta^\prime} \, \hat{A}_\mu^\prime \, e^{-i \delta^\prime} - i \bigl(\partial_\mu e^{i \delta^\prime}\bigr) \, e^{-i \delta^\prime},\\
	(\delta^\prime)^\dagger
	&= \delta^\prime
	= \delta^{\prime z} \, t_z.
\end{alignedeqn}
Infinitesimal gauge transformations of $A_\mu^\prime$ act for small $\delta^\prime$ as
\begin{equation}\label{eqn:214D}
	\delta^\prime
	\to \delta^\prime + \varphi + \frac{i}{2} [\varphi
	,\delta^\prime] + \dots
\end{equation}
The gauge-transformed $\delta^{\prime z}(\varphi)$,
\begin{equation}\label{eqn:214E}
	\delta^{\prime z}(\varphi)
	= \delta^{\prime z}(0) + \varphi^z - \tfrac{1}{2} \tensor{f}{_v_w^z} \, \varphi^v \, \delta^{\prime w}(0),
\end{equation}
obeys
\begin{equation}\label{eqn:214F}
	\tensor{N}{^z_y}
	= \frac{\partial \delta^{\prime z}}{\partial \varphi^y}
	= \delta_y^z - \tfrac{1}{2} \tensor{f}{_y_w^z} \, \delta^{\prime w}
	= \tensor{\bigl(1 - \tfrac{i}{2} \delta^\prime\bigr)}{^z_y}.
\end{equation}
The Faddeev-Popov determinant $\tilde{M} = \det N$, evaluated for $\delta^\prime = 0$ as appropriate for $\alpha \to 0$, simply becomes $\tilde{M} = 1$. Furthermore, this choice of the gauge fixing ensures the absence of Gribov copies since $\delta^\prime = 0$ guarantees that $\hat{A}_\mu$ is the unique representative of the gauge orbit.

Expanding eq.~\eqref{eqn:214C} for small $\delta^\prime$ yields
\begin{equation}\label{eqn:214H}
 	A_\mu^\prime - \hat{A}_\mu^\prime
 	= D_\mu(\hat{A}^\prime) \, \delta^\prime - \frac{i}{2} [D_\mu(\hat{A}^\prime) \, \delta^\prime,\delta^\prime] + \dots,
\end{equation}
The lowest order term replaces in eq.~\eqref{eqn:A15}, \eqref{eqn:191A} the macroscopic physical field $\hat{A}$ by the microscopic physical field $\hat{A}^\prime = \hat{A} + \hat{h}$. Higher orders can be obtained by an iterative solution of eq.~\eqref{eqn:214H}. It is tempting to consider a Landau-type of gauge fixing with $\hat{A}$ replaced by $\hat{A}^\prime$.

The gauge fixing based on eq.~\eqref{eqn:214C} is a physical gauge fixing in the sense that it only affects the gauge fluctuations around $\hat{A}^\prime$. Concerning the macroscopic field this does not imply $\hat{c} = 0$, however. Only physical microscopic fields $\hat{A}^\prime$ contribute effectively in the functional integral. However, a linear combination of two physical microscopic fields is typically not a physical field. As a result, the expectation value
\begin{equation}\label{eqn:OG1}
	\tilde{A}
	= \langle A^\prime\rangle
\end{equation}
will not be a macroscopic physical field $\hat{A}$ and we cannot infer $\hat{c} = 0$.

What is possible, however, is a unique association between a macroscopic physical field $\hat{A}$ and the expectation value $\tilde{A}$, such that $\tilde{A}$ is a gauge transform of $\hat{A}$,
\begin{alignedeqn}\label{eqn:OG2}
	&\tilde{A}_\mu
	= W \, \hat{A}_\mu \, W^\dagger - i (\partial_\mu W) \, W^\dagger,\\
	&W^\dagger \, W = 1,
	\qquad
	\det W
	= 1.
\end{alignedeqn}
The map $\hat{A}(\tilde{A})$ is defined here by the general map of every gauge field to a physical field. The existence of the inverse $\tilde{A}(\hat{A})$ is possible since for $\alpha \to 0$ the gauge fixing \eqref{eqn:A16}, \eqref{eqn:214C} results in a restriction for the possible values of $\tilde{A}$. Only those $\tilde{A}$ can be realized that are linear combinations of physical gauge fields. The manifolds spanned by the family of physical fields $\hat{A}$ and the one spanned by the expectation values $\tilde{A}$ therefore permit a one-to-one mapping if $\tilde{A}$ is the only possible expectation value on the gauge orbit of $\hat{A}$. To every $\hat{A}$ corresponds then precisely one $\tilde{A}$.

A unique map $\tilde{A}(\hat{A})$ allows us to define
\begin{equation}\label{eqn:OG3}
	\bar\Gamma[\hat{A}]
	= \check\Gamma[\tilde{A}(\hat{A})],
\end{equation}
where $\check\Gamma[\tilde{A}]$ is obtained from the usual gauge-fixed effective action $\tilde\Gamma[A]$ by inserting the partial solution of the field equations in the presence of the gauge-fixing term corresponding to eq.~\eqref{eqn:A16}, \eqref{eqn:214C}. This parallels the construction of $\bar\Gamma[\hat{A}]$ from $\tilde\Gamma[A]$ in sect.~\ref{sec:functional integral}. Then the gauge-invariant effective action is defined as before
\begin{equation}\label{eqn:OG4}
	\bar\Gamma[A]
	= \bar\Gamma[\hat{A}(A)]
	= \check\Gamma\bigl[\tilde{A}\bigl(\hat{A}(A)\bigr)\bigr]
	= \check\Gamma[\tilde{A}(A)].
\end{equation}
As it should be, the source
\begin{equation}\label{eqn:OG5}
	J
	= \frac{\partial \bar\Gamma}{\partial A}
	= \frac{\partial \check\Gamma}{\partial \tilde{A}} \frac{\partial \tilde{A}}{\partial A\vphantom{\tilde{A}}}
	= L \, \frac{\partial \tilde{A}}{\partial A}
	= L \, \frac{\partial \tilde{A}}{\partial \hat{A}} \frac{\partial \hat{A}}{\partial A\vphantom{\tilde{A}}}
\end{equation}
is covariantly conserved with respect to the macroscopic field $A$, $D_\mu(A) \, J^\mu = 0$, due to the factor $\partial \hat{A}/\partial A$.

This construction demonstrates that an optimal physical gauge fixing is possible. We leave the question of its practical use to future investigation.

\section{Expectation values for implicit definition of effective action}
\label{app:expectation values}

In this appendix we discuss the implicit definition of the effective action by a functional differential equation. We first present a general discussion, and subsequently specialize to Yang-Mills theories.

\subsection{Implicit definition of effective action}

For a general discussion we consider arbitrary ``fields'' or variables $\phi^\prime$, $\phi$. The microscopic variables $\phi^\prime$ and the macroscopic variables $\phi$ are vectors with components $\phi_i^\prime$ and $\phi_i$, which may be associated to fields such as $A_\mu^z(x)$, e.g. $i = (x,\mu,z)$. We define the effective action by the functional differential equation
\begin{equation}\label{eqn:EA1}
	\Gamma[\phi]
	= -\ln \int \mathcal{D} \phi^\prime \, \exp\Bigl\{-\tilde{S}[\phi^\prime,\phi] + \tfrac{\partial \Gamma}{\partial \phi} \, (\phi^\prime - \phi)\Bigr\},
\end{equation}
where $(\partial \Gamma/\partial \phi) \, (\phi^\prime - \phi)$ stands for the scalar product $(\partial \Gamma/\partial \phi_i) \, (\phi_i^\prime - \phi_i)$. If $\tilde{S}$ is independent of $\phi$ eq.~\eqref{eqn:EA1} is a standard relation for the effective action, as constructed by the Legendre transform of the generating functional $W$ for the connected correlation functions. We consider here the case where the action $\tilde{S}$ is also allowed to depend on the macroscopic field $\phi$. With
\begin{equation}\label{eqn:EA2}
	Z[\phi]
	= \int \mathcal{D} \phi^\prime \, \exp\Bigl\{-\tilde{S}[\phi^\prime,\phi] + \tfrac{\partial \Gamma}{\partial \phi} \, \phi^\prime\Bigr\},
\end{equation}
One has
\begin{equation}\label{eqn:EA3}
	\Gamma
	= -\ln Z + \frac{\partial \Gamma}{\partial \phi} \, \phi.
\end{equation}

Taking a derivative of eq.~\eqref{eqn:EA3} with respect to $\phi$ yields the identity
\begin{equation}\label{eqn:EA4}
	\Gamma^{(2)}(\langle\phi^\prime\rangle - \phi)
	= \biggl\langle\frac{\partial \tilde{S}}{\partial \phi}\biggr\rangle,
\end{equation}
with $\partial \tilde{S}/\partial \phi$ taken at fixed $\phi^\prime$. Here expectation values are defined as
\begin{equation}\label{eqn:EA5}
	\langle F[\phi^\prime,\phi]\rangle
	= \frac{1}{Z} \int \mathcal{D} \phi^\prime \, F[\phi^\prime,\phi] \exp\Bigl\{-\tilde{S}[\phi^\prime,\phi] + \tfrac{\partial \Gamma}{\partial \phi} \, \phi^\prime\Bigr\}.
\end{equation}
In general, expectation values depend on the sources $L$. The choice \eqref{eqn:EA1}, \eqref{eqn:EA3} identifies $L = \partial \Gamma/\partial \phi$. Eq.~\eqref{eqn:EA4} involves the matrix of second derivatives,
\begin{equation}\label{eqn:EA6}
	\Gamma_{ij}^{(2)}
	= \frac{\partial^2 \Gamma}{\partial \phi_i \, \partial \phi_j},
	\qquad
	\Gamma^{(2)} \, \phi
	= \Gamma_{ij}^{(2)} \, \phi_j.
\end{equation}

We can use the relation \eqref{eqn:EA4} in order to compute the expectation value of $\phi^\prime$ for a source given by $\partial \Gamma/\partial \phi$. Indeed, for $\tilde{S}$ independent of $\phi$ and invertible $\Gamma^{(2)}$ the solution of eq.~\eqref{eqn:EA4} identifies the macroscopic field with the expectation value of the microscopic field,
\begin{equation}\label{eqn:EA7}
	\phi
	= \langle\phi^\prime\rangle.
\end{equation}
On the other hand, for $\langle\partial \tilde{S}/\partial \phi\rangle \neq 0$ and finite invertible $\Gamma^{(2)}$ the expectation value $\langle\phi^\prime\rangle$ is no longer given by the macroscopic field $\phi$.

There are two cases for which $\langle\phi^\prime\rangle = \phi$ holds even in the case where $\tilde{S}$ depends on $\phi$. The first is simply that the expectation value of the $\phi$-derivative of $\tilde{S}$ vanishes or is proportional to $\langle\phi^\prime\rangle - \phi$, $\langle\partial \tilde{S}/\partial \phi\rangle = D \bigl(\langle\phi^\prime\rangle - \phi\bigr)$. The second corresponds to the case where $\Gamma^{(2)}$ contains a diverging piece, $\alpha \to 0$,
\begin{equation}\label{eqn:EA8}
	\Gamma^{(2)}
	= \frac{1}{\alpha} \, (1 - P) \, B \, (1 - P) + \dots
\end{equation}
with $P$ a projector. If the omitted terms denoted by the dots are finite for $\alpha \to 0$, the inverse of $\Gamma^{(2)}$ takes the form
\begin{equation}\label{eqn:EA9}
	G
	= \bigl(\Gamma^{(2)}\bigr)^{-1}
	= P \, C \, P + \dots,
\end{equation}
where the omitted terms corresponding to the dots vanish $\sim \alpha$. For
\begin{equation}\label{eqn:EA10}
	P \biggl\langle\frac{\partial \tilde{S}}{\partial \phi}\biggr\rangle
	= D \bigl(\langle\phi^\prime\rangle - \phi\bigr)
\end{equation}
one has $\langle\phi^\prime\rangle = \phi$.

\subsection{Yang-Mills theory}

We want to investigate for pure Yang-Mills theories with physical gauge fixing under which conditions eq.~\eqref{eqn:EA10} is obeyed, and therefore $\langle A_\mu^\prime\rangle = \hat{A}_\mu$. Gauge fixings with $\langle A_\mu^\prime\rangle = \hat{A}_\mu$ are called optimal. We will see that Landau gauge fixing is not an optimal gauge fixing. As discussed in app.~\ref{app:physical gauge fixing} the notion of optimal physical gauge fixing can be extended to the case where $\langle A_\mu^\prime\rangle$ is a unique representative in the gauge orbit of $\hat{A}_\mu$.

For gauge theories we may denote
\begin{equation}\label{eqn:EA11}
	V_z^\mu
	= \biggl\langle\frac{\partial \tilde{S}}{\partial A_\mu^z}\biggr\rangle
	= \biggl\langle\frac{\partial}{\partial A_\mu^z} \bigl(S_\text{gf} - \ln M\bigr)\biggr\rangle,
\end{equation}
with derivative taken at fixed $A^\prime$. Pieces in $V_z^\mu$ that are either longitudinal, e.g. $\Delta V_z^\mu = \tensor{(D_\mu)}{_z^y} \, \tilde\kappa_y$, or are proportional to $\langle A^\prime - \hat{A}\rangle$, do not contribute to the difference between $\langle A_\mu^{\prime z}\rangle$ and $\hat{A}_\mu^z$. For our choice of gauge fixing $S_\text{gf}$ and $\tilde{M}$ depend on $A$ only through their dependence on $\hat{A}$. With
\begin{equation}\label{eqn:GG1}
	\hat{V}_z^\mu
	= \frac{\partial \tilde{S}}{\partial \hat{A}_\mu^z}
\end{equation}
one has
\begin{equation}\label{eqn:GG2}
	V_z^\mu
	= \frac{\partial \hat{A}_\nu^y}{\partial A_\mu^z} \, \langle\hat{V}_y^\nu\rangle,
\end{equation}
such that $V_z^\mu$ is transversal due to the relation \eqref{eqn:CD}. On the other hand, any longitudinal component of $\hat{V}_z^\mu$ does not contribute to $\langle A^\prime\rangle - \hat{A}$, see eq.~\eqref{eqn:CE}.

The expectation value of $\partial S_\text{gf}/\partial A_\mu^z$ reads
\begin{equation}\label{eqn:EA12}
	\frac{\partial S_\text{gf}}{\partial \hat{A}_\mu^z}
	= \frac{1}{\alpha} \biggl\{\bigl(D^\mu D^\nu c_\nu^\prime\bigr)_z - f_{zyw}(D^\nu c_\nu^\prime)^y c^{\prime \mu w}\biggr\}.
\end{equation}
where $D^\mu = D^\mu(\hat{A})$ and we employ
\begin{equation}\label{eqn:EAW}
	\frac{\partial}{\partial \hat{A}_\mu^z} \, D_\rho
	= -i \delta_\rho^\mu \, T_z,
	\quad
	\frac{\partial}{\partial \hat{A}_\mu^z} \, \tensor{(D_\rho)}{^y_w}
	= -\delta_\rho^\mu \, \tensor{f}{_z^y_w}.
\end{equation}
Both $\langle c_\mu^{\prime z}(x)\rangle$ and $\langle c_\mu^{\prime z}(x) \, c_\nu^{\prime y}(y)\rangle$ vanish $\sim \alpha$. Indeed, the correlation function in the longitudinal sector, $\langle c^\prime \, c^\prime\rangle - \langle c^\prime\rangle \langle c^\prime\rangle \sim (1 - P) \bigl(\tilde\Gamma^{(2)}\bigr)^{-1} (1 - P)$, vanishes $\sim \alpha$. This cancels the factor $\alpha^{-1}$ in eq.~\eqref{eqn:EA12}, but is not sufficient for $\langle \partial S_\text{gf}/\partial A_\mu^z\rangle$ to vanish. The first term in eq.~\eqref{eqn:EA12} vanishes since it is longitudinal. For the second term we employ eq.~\eqref{eqn:G4},
\begin{equation}\label{eqn:EAY}
	c_\mu^{\prime z}
	= \tensor{(D_\mu)}{^z_y} \, \tilde{c}^y,
\end{equation}
such that
\begin{equation}\label{eqn:EAX}
	\frac{\partial S_\text{gf,2}}{\partial \hat{A}_\mu^z}
	= -\frac{1}{\alpha} \, f_{zvw} \, (D^\mu \tilde{c})^w \, (D^2 \tilde{c})^v.
\end{equation}
For the expectation value we use
\begin{equation}\label{eqn:EAN1}
	\langle\tilde{c}^y \, \tilde{c}^u\rangle
	= \langle\tilde{c}^y \, \tilde{c}^u\rangle_c
	= \alpha (D^{-4})^{yu}
\end{equation}
such that
\begin{alignedeqn}\label{eqn:EAN2}
	\biggl\langle\frac{\partial S_\text{gf,2}}{\partial \hat{A}_\mu^z}\biggr\rangle
	&= - f_{zvw} \tensor{(D^\mu)}{^w_y} (D^{-2})^{yv}\\
	&= -i \Tr\bigl\{T_z \, D^\mu \, D^{-2}\bigr\}.
\end{alignedeqn}
Indeed, eq.~\eqref{eqn:EAN1} follows from the longitudinal part of $\Gamma^{(2)}$ (cf. sect.~\ref{sec:gauge-invariant flow}),
\begin{alignedeqn}\label{eqn:EAN3}
	\bigl[(1 - P) \, &\Gamma^{(2)} \, (1 - P)\bigr]_{\mu y}^{z \nu}\\
	&= \frac{1}{\alpha} (\mathcal{D}_L)_{\mu y}^{z \nu}= -\frac{1}{\alpha} \tensor{(D_\mu)}{^z_w} \tensor{(D_\nu)}{^w_y},
\end{alignedeqn}
(for configurations where $(\mathcal{D}_L)_{\mu y}^{z \nu}$ is symmetric), which is inverted by
\begin{equation}\label{eqn:EAN4}
	\langle c_\mu^{\prime z} \, c_\nu^{\prime y}\rangle
	= -\alpha D_\mu D^{-4} D_\nu.
\end{equation}
With $c_\mu^{\prime z} = \tensor{(D^\mu)}{^z_y} \, \tilde{c}^y$ and taking into account the antisymmetry of $D_\mu$ this is equivalent to eq.~\eqref{eqn:EAN1}. (One may equivalently use directly the effective action for $\tilde{c}$.)

For the contribution from the Faddeev-Popov determinant one employs
\begin{alignedeqn}\label{eqn:EA13}
	-\frac{\partial}{\partial \hat{A}_\mu^z} \, \ln M
	&= -\tensor{f}{_z_v^y} \, \tensor{\bigl(D^\mu(A^\prime)\bigr)}{_y^w} \, \tensor{(G_\text{gh})}{_w^v}\\
	&= -i \Tr\bigl\{T_z \, D^\mu(A^\prime) \, G_\text{gh}\bigr\},
\end{alignedeqn}
with ghost propagator
\begin{equation}\label{eqn:EA14}
	G_\text{gh}
	= \bigl[-D_\mu(\hat{A}) \, D^\mu(A^\prime)\bigr]^{-1}.
\end{equation}
Combining eq.~\eqref{eqn:EAX}, \eqref{eqn:EA13} one finds
\begin{alignedeqn}\label{eqn:HH1}
	\hat{V}_z^\mu
	= -i \Tr\bigl\{&T_z \, D^\mu(A^\prime) \, G_\text{gh}(A^\prime,\hat{A})\\
	&- D^\mu(\hat{A}) \, G_\text{gh}(\hat{A},\hat{A})\bigr\},
\end{alignedeqn}
such that $\hat{V}_z^\mu$ vanishes for $A^\prime = \hat{A}$. Expanding $\hat{V}_z^\mu$ in $A^\prime - \hat{A}$, the linear term in $A^\prime - \hat{A}$ does not contribute to $\langle A^\prime\rangle - \hat{A}$. Higher-order terms in the expansion contribute, however, to $\langle\hat{V}_z^\mu\rangle$ even for $\langle A^\prime\rangle = \hat{A}$.

We conclude that $\langle\hat{V}_z^\mu\rangle$ does, in general, not vanish for $\langle A^\prime\rangle = \hat{A}$. From eq.~\eqref{eqn:EA4} we infer a difference between $\langle A^\prime\rangle$ and $\hat{A}$ according to
\begin{equation}\label{eqn:HH2}
	\langle A_\mu^{\prime z}\rangle - \hat{A}_\mu^z
	= (G_P)_{\mu y}^{z \nu} \, V_\nu^y.
\end{equation}

\section{Correlation function for implicit definition of effective action}
\label{app:correlation function}

In this appendix we use the functional differential equation \eqref{eqn:96B} or \eqref{eqn:EA1} in order to establish a general expression for the two-point correlation function. For pure gauge theories and optimal physical gauge fixing the correlation function equals the propagator, establishing eq.~\eqref{eqn:CF6}.

In order to obtain a relation for the second functional derivative of $\Gamma$ we start from
\begin{alignedeqn}\label{eqn:MN1}
	&\Gamma
	= -\ln \tilde{Z},
	\qquad
	\tilde{Z}
	= \int \mathcal{D} \phi^\prime \, \exp(-\hat{S}),\\
	&\hat{S}
	= \tilde{S} - \frac{\partial \Gamma}{\partial \phi_l} \, (\phi_l^\prime - \phi_l).
\end{alignedeqn}
Derivatives with respect to $\phi$ yield
\begin{equation}\label{eqn:MN2}
	\frac{\partial \Gamma}{\partial \phi_i}
	= \biggl\langle\frac{\partial \hat{S}}{\partial \phi_i}\biggr\rangle
\end{equation}
and
\begin{equation}\label{eqn:MN3}
	\frac{\partial^2 \Gamma}{\partial \phi_i \, \partial \phi_j}
	= \frac{\partial \Gamma}{\partial \phi_i} \frac{\partial \Gamma}{\partial \phi_j} + \biggl\langle\frac{\partial^2 \hat{S}}{\partial \phi_i \, \partial \phi_j} - \frac{\partial \hat{S}}{\partial \phi_i} \frac{\partial \hat{S}}{\partial \phi_j}\biggr\rangle
\end{equation}
with expectation value defined by
\begin{equation}\label{eqn:MN4}
	\langle B\rangle
	= \tilde{Z}^{-1} \int \mathcal{D} \phi^\prime \, B \, \exp(-\hat{S}).
\end{equation}
We next exploit
\begin{equation}\label{eqn:MN5}
	\frac{\partial \hat{S}}{\partial \phi_i}
	= \frac{\partial \Gamma}{\partial \phi_i} + \frac{\partial \tilde{S}}{\partial \phi_i} - \frac{\partial^2 \Gamma}{\partial \phi_i \, \partial \phi_l} \, (\phi_l^\prime - \phi_l)
\end{equation}
and
\begin{equation}\label{eqn:MN6}
	\frac{\partial^2 \hat{S}}{\partial \phi_i \, \partial \phi_j}
	= 2 \, \frac{\partial^2 \Gamma}{\partial \phi_i \, \partial \phi_j} + \frac{\partial^2 \tilde{S}}{\partial \phi_i \, \partial \phi_j} - \frac{\partial^3 \Gamma}{\partial \phi_i \, \partial \phi_j \, \partial \phi_l} \, (\phi_l^\prime - \phi_l).
\end{equation}
Insertion into eq.~\eqref{eqn:MN3} and use of eq.~\eqref{eqn:EA4} yields
\begin{equation}\label{eqn:MN7}
	\frac{\partial^2 \Gamma}{\partial \phi_j \, \partial \phi_m} \, \frac{\partial^2 \Gamma}{\partial \phi_i \, \partial \phi_l} \langle(\phi_l^\prime - \phi_l) (\phi_m^\prime - \phi_m)\rangle
	= \frac{\partial^2 \Gamma}{\partial \phi_i \, \partial \phi_j} + X_{ij}
\end{equation}
with
\begin{align}\label{eqn:MN8}
	X_{ij}
	&= \biggl\langle\frac{\partial^2 \tilde{S}}{\partial \phi_i \, \partial \phi_j} - \frac{\partial \tilde{S}}{\partial \phi_i} \frac{\partial \tilde{S}}{\partial \phi_j}\biggr\rangle\\
	&\hphantom{{}=}+ \biggl\langle\!\biggl(2 \, \frac{\partial^2 \Gamma}{\partial \phi_j \, \partial \phi_l} \frac{\partial \tilde{S}}{\partial \phi_i} - \frac{\partial^3 \Gamma}{\partial \phi_i \, \partial \phi_j \, \partial \phi_l}\biggr) (\phi_l^\prime - \phi_l)\biggr\rangle.\notag
\end{align}
(The expression for $X$ should be symmetrized in $i$ and $j$.) For $\tilde{S}$ independent of $\phi$ one has $\langle \phi_l^\prime\rangle = \phi_l$ such that $X$ vanishes.

We next multiply eq.~\eqref{eqn:MN7} with $\bigl(\Gamma^{(2)}\bigr)^{-1}$, such that
\begin{equation}\label{eqn:MN9}
	\bigl(\Gamma^{(2)}\bigr)_{il} \, \tilde{G}_{lm}
	= \delta_{im} + Y_{im},
\end{equation}
with
\begin{equation}\label{eqn:MN10}
	\tilde{G}_{lm}
	= \langle(\phi_l^\prime - \phi_l) (\phi_m^\prime - \phi_m)\rangle
\end{equation}
and
\begin{equation}\label{eqn:MN11}
	Y_{im}
	= X_{ij} \, \bigl(\Gamma^{(2)}\bigr)_{jm}^{-1}.
\end{equation}

For $Y = 0$ one recovers the well-known expression of the propagator $\tilde{G}$ as the inverse of the second functional derivative of $\Gamma$. More generally, one has
\begin{alignedeqn}\label{eqn:MN12}
	&\tilde{G}_{ij}
	= G_{ij} + \Delta G_{ij},
	\qquad
	G_{ij}
	= \bigl(\Gamma^{(2)}\bigr)_{ij}^{-1},\\
	&\Delta G_{ij}
	= G_{il} \, X_{lm} \, G_{mj}.
\end{alignedeqn}
The ``correction term'' can be written as
\begin{equation}\label{eqn:MN13}
	\Delta G_{ij}
	= \Delta G_{ij}^{(1)} + \Delta G_{ij}^{(2)} + \Delta G_{ij}^{(3)},
\end{equation}
where
\begin{alignedeqn}\label{eqn:MN14}
	&\Delta G_{ij}^{(1)}
	= G_{il} \biggl\langle\frac{\partial^2 \tilde{S}}{\partial \phi_l \, \partial \phi_m} - \frac{\partial \tilde{S}}{\partial \phi_l} \frac{\partial \tilde{S}}{\partial \phi_m}\biggr\rangle G_{mj},\\
	&\Delta G_{ij}^{(2)}
	= 2 \biggl\langle\frac{\partial \tilde{S}}{\partial \phi_m} \, (\phi_i^\prime - \phi_i)\biggr\rangle G_{mj},\\
	&\Delta G_{ij}^{(3)}
	= \frac{\partial G_{ij}}{\partial \phi_l} \langle\phi_l^\prime - \phi_l\rangle.
\end{alignedeqn}

\section{Gauge-invariant flow equation from implicit definition of effective action}
\label{app:gauge-invariance flow}

In this appendix we show that an implicit definition of the average effective action by a functional differential equation leads to an exact closed flow equation. It is formulated here for Landau gauge fixing. This flow equation is not yet optimal as the one proposed in ref.~\cite{CWGIF}. We briefly discuss which modifications can lead to the flow equation \eqref{eqn:I1}.

We start by adding an infrared cutoff term $\Delta S_k$ in the implicit definition \eqref{eqn:96B} of the effective action
\begin{align}\label{eqn:FE1}
	&\tilde\Gamma_k[A]
	= -\ln \int \mathcal{D} A^\prime \, E_k(A)\\
	&\times \exp\biggl\{-(\tilde{S} + \Delta S_k)[A^\prime,A] + \int_x \frac{\partial \tilde\Gamma_k}{\partial A_\mu^z} \, \bigl(A_\mu^{\prime z} - A_\mu^z\bigr)\biggr\},\notag
\end{align}
where $\tilde{S}$ contains a Landau gauge-fixing term and the corresponding logarithm of the Faddeev-Popov determinant, $\tilde{S} = S^\prime + S_\text{gf} - \ln M$. For the IR cutoff we take
\begin{equation}\label{eqn:FE2}
	\Delta S_k
	= \frac{1}{2} \int_x b_z^{\prime \mu} \bigl(R_k^\text{(ph)}\bigr)_{\mu y}^{z \nu} \, b_\nu^{\prime y} + \frac{1}{2 \alpha} \int_x c_z^{\prime \mu} \bigl(R_k^\text{(g)}\bigr)_{\mu y}^{z \nu} \, c_\nu^{\prime y},
\end{equation}
with $R_k^\text{(ph)}$ and $R_k^\text{(g)}$ suitable functions (see below) of covariant derivatives involving $\hat{A}(A)$. The factor $E_k(A)$ regularizes the Faddeev-Popov determinant (see below). It is a gauge-invariant expression of the covariant Laplacian $D^2(A)$.

Taking a logarithmic $k$-derivative yields, with $t = \ln(k/k_0)$
\begin{equation}\label{eqn:FE3}
	\partial_t \tilde\Gamma_k
	= \langle\partial_t \Delta S_k\rangle - \epsilon_k + \int_x \partial_t \frac{\partial \tilde\Gamma_k}{\partial A_\mu^z} \, \langle A_\mu^{\prime z} - A_\mu^z\rangle,
\end{equation}
with
\begin{equation}\label{eqn:FE4}
	\epsilon_k
	= \partial_t \ln E_k.
\end{equation}
We define the effective average action as
\begin{align}\label{eqn:FE5}
	\Gamma_k
	&= \tilde\Gamma_k - \frac{1}{2} b_z^\mu \bigl(R_k^\text{(ph)}\bigr)_{\mu y}^{z \nu} \, b_\nu^y - \frac{1}{2 \alpha} c_z^\mu \bigl(R_k^\text{(g)}\bigr)_{\mu y}^{z \nu} \, c_\nu^y\notag\\
	&\hphantom{{}=}- \int_x \frac{\partial \tilde\Gamma_k}{\partial A_\mu^z} (b_\mu^z + c_\mu^z - \hat{c}_\mu^z),
\end{align}
where $b = \langle b^\prime\rangle$, $c = \langle c^\prime\rangle$. It obeys the flow equation (at fixed $A$)
\begin{alignedeqn}\label{eqn:FE6}
	&\partial_t \Gamma_k
	= \pi_k + \delta_k - \epsilon_k,\\
	&\pi_k
	= \frac{1}{2} \int_x \partial_t \bigl(R_k^\text{(ph)}\bigr)_{\mu y}^{z \nu} \, \langle b_\nu^{\prime y} b_z^{\prime \mu}\rangle_c,\\
	&\delta_k
	= \frac{1}{2 \alpha} \int_x \partial_t \bigl(R_k^\text{(g)}\bigr)_{\mu y}^{z \nu} \, \langle c_\nu^{\prime y} c_z^{\prime \mu}\rangle_c,
\end{alignedeqn}
with $\langle\cdot\rangle_c$ denoting the connected two-point function. We recognize the correlation function for the physical fluctuations
\begin{equation}\label{eqn:FE7}
	\langle b_\nu^{\prime y} b_z^{\prime \mu}\rangle_c
	= \bigl(\tilde{G}_P\bigr)_{\nu z}^{y \mu}
\end{equation}
obeying
\begin{equation}\label{eqn:FE8}
	P \, \tilde{G}_P
	= \tilde{G}_P \, P^\transp
	= \tilde{G}_P.
\end{equation}
The correlation function for the gauge fluctuations,
\begin{equation}\label{eqn:FE9}
	\langle c_\nu^{\prime y} c_z^{\prime \mu}\rangle_c
	= \bigl(\tilde{G}_g\bigr)_{\nu z}^{y \mu}
\end{equation}
vanishes $\sim \alpha$. Its leading expression in a saddle point expansion obeys
\begin{equation}\label{eqn:FE10}
	\frac{1}{\alpha} \bigl\{-\tensor{\bigl(D_\rho \, D^\nu\bigr)}{^w_y} + \bigl(R_k^\text{(g)}\bigr)_{\rho y}^{w \nu}\bigr\} \bigl(\tilde{G}_g\bigr)_{\nu z}^{y \mu}
	= (1 - P)_{\rho z}^{w \mu},
\end{equation}
such that
\begin{alignedeqn}\label{eqn:FE11}
	&\delta_k
	= \frac{1}{2} \tr\bigl\{\partial_t R_k^\text{(g)} \, \bigl(\mathcal{D}_L + R_k^\text{(g)}\bigr)^{-1}\bigr\},\\
	&\tensor{(\mathcal{D}_L)}{_\rho^\nu}
	= -D_\rho D^\nu.
\end{alignedeqn}
Typically, $R_k^\text{(g)}$ is chosen as a function of $\mathcal{D}_L$. We will find $\epsilon_k = -2 \delta_k$, such that eq.~\eqref{eqn:FE6} yields eq.~\eqref{eqn:I1}, if $\tilde{G}_P = G_P$, with $G_P$ obeying eq.~\eqref{eqn:I6} with $R_P$ identified with $R_k^\text{(ph)}$.

For $k \neq 0$ the dependence of the cutoff function on the macroscopic gauge field $A$, $\partial R_k^\text{(ph)}/\partial A_\mu^z \neq 0$ is responsible for additional contributions to $b_\mu^z = \langle b_\mu^{\prime z}\rangle \neq 0$, even if for an optimal cutoff one has $b = 0$ for $k = 0$. The macroscopic gauge field $A_\mu^z$ does not equal the expectation value of the microscopic gauge field. The addition of $\Delta S_k$ does not change the discussion in the gauge fixing and Faddeev-Popov sectors, such that one still has
\begin{equation}\label{eqn:FE12}
	\langle c^\prime\rangle
	= c
	= \hat{c}
	= 0.
\end{equation}
However, for $b_\mu^z \neq 0$ one has now
\begin{equation}
	\langle A_\mu^{\prime z}\rangle
	= A_\mu^z + b_\mu^z.
\end{equation}
For the definition \eqref{eqn:FE5} of $\Gamma_k$ it seems likely that the difference $\Delta G = \tilde{G} - G$, computed according to eq.~\eqref{eqn:MN12}, does not vanish. This does not prevent the flow equation for the gauge-invariant effective action to be a gauge-invariant closed equation. The insertion of the ``correction term'' $\Delta G$ remains computable in terms of $\Gamma$, with $\tilde{S}$ in eq.~\eqref{eqn:MN8} modified by the addition of the physical part of $\Delta S_k$. Correction terms are proportional to $\partial R_k^\text{(ph)}/\partial A_\mu^z$ or derivatives of it, and vanish for $k \to 0$.

In order to obtain a simpler flow equation were corrections from $\Delta G$ are absent we may modify the precise definition of the effective average action $\Gamma_k$ for $k \neq 0$. This influences the relation between the expectation value $\langle A_\mu^{\prime z}\rangle$ and the macroscopic field $A_\mu^z$, as well as $\Delta G$. Also the relation between the source $L$ and the macroscopic gauge field $A$ may be modified, such that $L$ is no longer given by $\partial \Gamma/\partial \phi$. If we do not fix the relation between $L$ and $\partial \Gamma/\partial \phi$ a priori, the relation between $\phi$ and $\langle \phi^\prime\rangle$ can be freely chosen, determining a posteriori the relation between $L$ and $\partial \Gamma/\partial \phi$. More formally, we may replace on the r.h.s. of eq.~\eqref{eqn:FE1} $\partial \tilde\Gamma/\partial A$ by some suitably chosen source functional $L[A]$, and add to the definition of $\Gamma_k$ in eq.~\eqref{eqn:FE5} a gauge-invariant term $\tilde{C}_k[A]$. We require that for $k = 0$ one has $L[A] = \partial \tilde\Gamma/\partial A$ and $\tilde{C}_k[A] = C[A]$. The correction $\Delta G$ depends on the choice of $L[A]$ and $\tilde{C}_k[A]$.

It has been argued in ref.~\cite{CWGIF} that a suitable definition of $\Gamma_k$ and the relation between $\phi$ and $\langle \phi^\prime\rangle$ exists such that the correction term $\sim \Delta G$ vanishes. In this paper we assume that this is indeed possible and work with the flow equation \eqref{eqn:I1}. We emphasize that different definitions of $\Gamma_k$, which lead to the same effective action for $k = 0$, yield the same expectation values and correlations for physical observables.

\end{appendices}





\bibliography{Gauge-Invariant_Fields}

\end{document}